\documentclass[preprint,10pt]{elsarticle}

\usepackage{amssymb}

\journal{Physics Letters B}
\usepackage[colorlinks=true,pdfstartview=FitV,linkcolor=blue,citecolor=blue,urlcolor=blue]{hyperref}
\usepackage[usenames,dvipsnames]{color}
\usepackage[sc]{mathpazo}
\usepackage[below]{placeins}
\usepackage{afterpage}
\usepackage[separate-uncertainty,retain-explicit-plus,per-mode = symbol]{siunitx}
\usepackage{booktabs}
\usepackage{longtable}
\usepackage{multirow}
\usepackage{rotating}
\usepackage{lineno}
\usepackage{graphicx}
\usepackage{bm}
\usepackage{paralist}
\usepackage{wrapfig}
\usepackage{appendix}
\usepackage{vruler}
\usepackage{hyperref}
\usepackage{etoolbox}
\usepackage{tikz}
\usepackage{listings}
\newcolumntype{d}[1]{D{.}{\cdot}{#1}}
\newcommand{\myrefs}[2]{\href{http://dx.doi.org/#2}{#1}}
\newcommand{\mref}[1]{\href{http://#1}{#1}}

\newcommand{\arxiv}[1]{\href{http://arxiv.org/abs/#1}{arxiv:#1}}

\DeclareSIUnit\c{$c$}
\DeclareSIUnit\week{w}
\DeclareSIUnit\year{yr}
\DeclareSIUnit\standard{std}
\DeclareSIUnit\str{sr}
\DeclareSIUnit\pe{PE}
\DeclareSIUnit\spe{SPE}
\DeclareSIUnit\ev{events}
\DeclareSIUnit\bin{(5-PE bin)}
\DeclareSIUnit\hit{hits}
\DeclareSIUnit\sgm{$\sigma$}
\DeclareSIUnit\rms{RMS}
\DeclareSIUnit\keVr{keV$_{\rm r}$}
\DeclareSIUnit\keVee{keV$_{\rm $e$e}$}
\DeclareSIUnit\ph{photons}
\DeclareSIUnit\neu{neutrons}
\DeclareSIUnit\pm{PMT}
\DeclareSIUnit\inch{''}
\DeclareSIUnit\bit{bit}
\DeclareSIUnit\sample{samples}
\DeclareSIUnit\barn{b}

\newcommand{\lith}{\mbox{$^{7}$Li}}
\newcommand{\lithexc}{\mbox{$^{7}$Li$^{*}$}}

\newcommand{\borten}{\mbox{$^{10}$B}}

\newcommand{\cfor}{\mbox{$^{14}$C}}

\newcommand{\ar}{\mbox{$^{39}$Ar}}

\newcommand{\coba}{\mbox{$^{60}$Co}}
\newcommand{\rbthree}{\mbox{$^{83}$Rb}}    
\newcommand{\rbthreetau}{\SI{124.4}{\day}}
\newcommand{\krthree}{\mbox{$^{83m}$Kr}}
\newcommand{\krthreetau}{\SI{2.64}{\hour}}
\newcommand{\krthreeinttau}{\SI{222}{\ns}}
\newcommand{\krthreefirstene}{\mbox{32.1\,keV}}
\newcommand{\krthreesecondene}{\mbox{9.4\,keV}}
\newcommand{\krthreepeakene}{\mbox{41.5\,keV}}
\newcommand{\tmbchem}{\mbox{B(OCH$_3$)$_3$}}

\newcommand{\ot}{\mbox{O$_2$}}

\newcommand{\nt}{\mbox{N$_2$}}

\newcommand{\ds}{\mbox{DarkSide}}
\newcommand{\dst}{\mbox{DarkSide-10}}
\newcommand{\dstexp}{502 days}								
\newcommand{\dsf}{\mbox{DarkSide-50}}

\newcommand{\bx}{\mbox{Borexino}}
\newcommand{\scene}{\mbox{SCENE}}
\newcommand{\scenenullfieldlyjun}{\SI{6.3(3)}{\pe\per\keV}}
\newcommand{\scenenullfieldlyoct}{\SI{4.8(2)}{\pe\per\keV}}

\newcommand{\lsv}{\mbox{LSV}}
\newcommand{\wcd}{\mbox{WCD}}

\newcommand{\tpc}{\mbox{TPC}}
\newcommand{\pmt}{\mbox{PMT}}

\newcommand{\tmb}{\mbox{TMB}}
\newcommand{\pc}{\mbox{PC}}

\newcommand{\ito}{\mbox{ITO}}
\newcommand{\ptfe}{\mbox{PTFE}}
\newcommand{\tpb}{\mbox{TPB}}

\newcommand{\lar}{\mbox{LAr}}

\newcommand{\uar}{\mbox{UAr}}

\newcommand{\sone}{\mbox{S1}}
\newcommand{\stwo}{\mbox{S2}}

\newcommand{\sthree}{\mbox{S3}}
\newcommand{\fno}{\mbox{f$_{90}$}}
\newcommand{\isovalgreen}{\SI{0.01}{\ev\per\bin}}

\newcommand{\bg}{\mbox{$\beta/\gamma$}}
\newcommand{\gr}{\mbox{$\gamma$-ray}}
\newcommand{\grs}{\mbox{$\gamma$-rays}}

\newcommand{\vesc}{\mbox{$v_{\rm escape}$}}
\newcommand{\vnaught}{\mbox{$v_0$}}
\newcommand{\vearth}{\mbox{$v_{\rm Earth}$}}
\newcommand{\rhodm}{\mbox{$\rho_{\rm dm}$}}
\newcommand{\laserspectrummean}{\mbox{$\mu_q$}}
\newcommand{\laserpedmean}{\mbox{$\mu_{\rm ped}$}}
\newcommand{\laserspemean}{\mbox{$\mu_{\si{\spe}}$}}
\newcommand{\laserpemean}{\mbox{$\mu_{\si{\pe}}$}}

\newcommand{\arwave}{\SI{128}{\nano\metre}}

\newcommand{\dsfnullfieldly}{\SI{7.9(4)}{\pe\per\keV}}   
\newcommand{\dsfdriftfieldly}{\SI{7.0(3)}{\pe\per\keV}}  

\newcommand{\dsflystability}{\SI{0.7}{\percent}}

\newcommand{\dsftriggereneth}{\SI{0.6}{\pe}}  

\newcommand{\dsfpmt}{{Hamamatsu R11065}}

\newcommand{\dsfruntimeneuexp}{$\sim$1.3}           

\newcommand{\dsfruntimeneuobs}{\num{4}}
\newcommand{\dsfruntimeneuradobs}{\num{3}}

\newcommand{\dsfbckneucos}{$\ll$1}          

\newcommand{\dsfbckargreen}{\SI{0.1}{\ev}}
\newcommand{\dsfactivevolumediameter}{35.6\,cm}
\newcommand{\dsfactivevolumeheight}{35.6\,cm}
\newcommand{\dsfactivemass}{\SI{46.4(7)}{\kg}}
\newcommand{\dsfuareqexponoerr}{\SI{215000}{\kg\day}}
\newcommand{\dsfuareqexpoty}{\SI{0.6}{\tonne\year}}
\newcommand{\gpsabsoluteaccuracy}{\SI{20}{\ns}}
\newcommand{\dsflimit}{\SI{6.1E-44}{\square\centi\meter}}
\newcommand{\dsflimitmass}{\SI{100}{\GeV\per\square\c}}

\newcommand{\dsfdriftfield}{\SI{200}{\volt\per\centi\meter}}
\newcommand{\dsfextractionfield}{\SI{2.8}{\kilo\volt\per\centi\meter}}
\newcommand{\dsfmultfield}{\SI{4.2}{\kilo\volt\per\centi\meter}}
\newcommand{\dsfcathodepot}{\SI{-12.7}{\kilo\volt}}
\newcommand{\dsfgridpot}{\SI{-5.6}{\kilo\volt}}

\newcommand{\dsftdriftmaxanal}{\SI{373}{\us}} 
\newcommand{\dsfelectronmeanlifefirst}{\SI{>3.5}{\ms}}
\newcommand{\dsfelectronmeanlifesecond}{\SI{>5}{ms}}

\newcommand{\dsfgaspocketthickness}{\SI{1}{\centi\meter}}
\newcommand{\dsflarovergrid}{\SI{5}{\milli\meter}}
\newcommand{\dsfpmtsize}{\SI{3}{\inch}}
\newcommand{\dsfpmtnum}{\num{38}}

\newcommand{\dsfpmtqe}{\SI{34}{\percent}}						

\newcommand{\dsfpmtwave}{\SI{420}{\nano\meter}}

\newcommand{\dsfcryoflowrate}{\SI{30}{\standard\liter\per\minute}}

\newcommand{\dsfcryopressure}{\SI{1080.0}{\milli\bar}}
\newcommand{\dsfcryopressurestability}{\SI{\pm0.1}{\milli\bar}}
\newcommand{\dsfcryorntrapartemp}{\SIrange[range-units = single]{185}{190}{\kelvin}}		

\newcommand{\dsfpmtgain}{\num{4.E5}}
\newcommand{\dsfitothickness}{\SI{15}{\nano\meter}}

\newcommand{\dsfthickptfereplector}{\SI{2.54}{\centi\meter}}
\newcommand{\dsftpbcenterthickness}{\SI{230(10)}{\micro\gram\per\square\centi\meter}}   
\newcommand{\dsftpbedgethickness}{\SI{190(15)}{\micro\gram\per\square\centi\meter}}     
\newcommand{\dsftpbwallhalfthickness}{\SI{165(20)}{\micro\gram\per\square\centi\meter}}   
\newcommand{\dsftpbwallbottomthickness}{\SI{224(27)}{\micro\gram\per\square\centi\meter}} 
\newcommand{\dsfgridfoil}{\SI{50}{\micro\meter}}
\newcommand{\dsfgridfoiltrasparency}{\SI{95}{\percent}}
\newcommand{\dsfcoldampgain}{\SI{3}{\volt/\volt}}
\newcommand{\dsfcoldampbandwith}{\SI{150}{\mega\hertz}}

\newcommand{\dsfcoldamplowfreqtimeconst}{\SI{5.8}{\milli\second}}
\newcommand{\dsfcoldamppeaktopeak}{\SI{3}{\volt}}
\newcommand{\dsfcoldamptermination}{\SI{50}{\ohm}}                               
\newcommand{\dsfcoldampnoise}{\SI{45}{\micro\volt}}
\newcommand{\dsfdigitresolution}{\SI{12}{\bit}}
\newcommand{\dsfdigitsamplespeed}{\SI{250}{\MHz}}
\newcommand{\dsffemgain}{\num{10}}
\newcommand{\dsftriggerwin}{100\,ns}
\newcommand{\dsfstwooversthree}{$\sim$\num{1000}}				
\newcommand{\dsftrigeff}{$>$\SI{99}{\percent}}
\newcommand{\dsftrigeffthresh}{\SI{60}{\pe}}
\newcommand{\dsftrigrate}{\SI{50}{\hertz}}
\newcommand{\dsftrigrategtwo}{\SI{13}{\hertz}}
\newcommand{\dsfacquiwindow}{\SI{440}{\us}}
\newcommand{\dsfinhibitwindow}{\SI{810}{\us}}
\newcommand{\dsfkrthreeobsrate}{\SIrange[range-units = single]{2}{3}{\hertz}}
\newcommand{\odpmtnum}{\num{190}}

\newcommand{\odfemgain}{\num{10}}
\newcommand{\oddigitresolution}{\SI{10}{\bit}}						
\newcommand{\oddigitsamplespeed}{\SI{1.25}{\GHz}}				
\newcommand{\odacquisitionwin}{\SI{70}{\us}}						

\newcommand{\odzerosuppressionrun}{\num{0.25}}					

\newcommand{\odneutronrejectionachieved}{\numrange{40}{60}}    
\newcommand{\odneutronrejectiondesign}{\num{200}}
\newcommand{\lsvdiameter}{\SI{4.0}{\m}}

\newcommand{\lsvpmtnum}{\num{110}}

\newcommand{\lsvpmt}{\mbox{Hamamatsu R5912}}
\newcommand{\lsvpmtsize}{\SI{8}{\inch}}
\newcommand{\lsvpmtqe}{\SI{37}{\percent}}
\newcommand{\lsvpmtwave}{\SI{408}{\nano\meter}}

\newcommand{\lsvly}{\SI{0.54(4)}{\pe\per\keV}}          
\newcommand{\lsvcforrate}{$\sim$\SI{150}{\kHz}}      

\newcommand{\lsvalphaequivenergy}{\SIrange[range-units = single]{50}{60}{\keV}}
\newcommand{\lsvscintillatormass}{\SI{30}{\tonne}}
\newcommand{\lsvppoconcentration}{\SI{2.5}{\gram\per\liter}}   
\newcommand{\lsvbtenxsec}{\SI{3840}{\barn}}
\newcommand{\lsvtmbzeropercentcapturetime}{\SI{250}{\micro\second}}
\newcommand{\lsvtmbfiftypercentcapturetime}{\SI{2.2}{\micro\second}}

\newcommand{\ctfpmtnum}{\num{80}}

\newcommand{\ctfpmt}{\mbox{ETL 9351}}
\newcommand{\ctfpmtsize}{\SI{8}{\inch}}
\newcommand{\ctfpmtqe}{\SI{27}{\percent}}
\newcommand{\ctfpmtwave}{\SI{420}{\nano\meter}}

\newcommand{\ctfheight}{\SI{10}{\meter}}
\newcommand{\ctfdiameter}{\SI{11}{\meter}}
\newcommand{\ctfwatermass}{\SI{1}{\kilo\tonne}}

\newcommand{\dsflaserpmtoccupancy}{\SI{0.1}{\pe}}
\newcommand{\dsflaserpulsewidth}{\SI{60}{\ps}}     
\newcommand{\dsflaserwavelength}{\SI{405}{nm}}   
\newcommand{\dsflaserpulserate}{\SI{500}{\hertz}}
\newcommand{\dsflaseracquiwindow}{\SI{3}{\micro\second}}
\newcommand{\dsflaserintegrationwindow}{\SI{108}{\ns}}
\newcommand{\dsflaserspemeanstability}{\SI{2.5}{\percent}}
\newcommand{\odlaserpulserate}{\SI{500}{\hertz}}

\newcommand{\movingaveragewindow}{\SI{80}{\ns}}
\newcommand{\sumchzerosuppthresh}{\SI{0.1}{\pe}}

\newcommand{\pulsestartthresh}{\SI{0.3}{\pe}}
\newcommand{\fixedintone}{\SI{7}{\micro\second}}
\newcommand{\fixedinttwo}{\SI{30}{\micro\second}}
\newcommand{\odpresamps}{\SI{20}{\sample}}
\newcommand{\odpresampstime}{\SI{16}{\ns}}
\newcommand{\odbaselinewindow}{\SI{15}{\sample}}
\newcommand{\sonezvariation}{\SI{19}{\percent}}   

\newcommand{\stwozvariation}{\mbox{7\%}}
\newcommand{\abborten}{\SI{20}{\percent}}
\newcommand{\enbortenexcitedgamma}{\SI{478}{\keV}}
\newcommand{\brbortenground}{\SI{6.4}{\percent}}
\newcommand{\enbortengroundalpha}{\SI{1775}{\keV}}
\newcommand{\brbortenexcited}{\SI{93.6}{\percent}}
\newcommand{\enbortenexcitedalpha}{\SI{1471}{\keV}}

\newcommand{\epar}{\SI{565}{\keV}}

\newcommand{\activityar}{\SI{1}{\becquerel\per\kg}}
\newcommand{\timefno}{\SI{90}{\nano\second}}
\newcommand{\longlivedtriplestatedimerargon}{$\sim$\SI{1.5}{\micro\second}}
\newcommand{\shortlivedtriplestatedimerargon}{\SI{6}{\nano\second}}

\newcommand{\uardepletion}{\num{150}}

\newcommand{\lngsequivdepth}{3800\,m.w.e.}  

\newcommand{\totalcalendartime}{\mbox{7 months}}

\newcommand{\dgcritliv}{\SI{53.8(2)}{\day}}
\newcommand{\dccritliv}{\SI{51.1(2)}{\day}}
\newcommand{\srcritliv}{\SI{49.2(2)}{\day}}

\newcommand{\nccutliv}{\SI{48.8(2)}{\day}}

\newcommand{\bacutliv}{\SI{48.8(2)}{\day}}

\newcommand{\licutmin}{\SI{1.35}{\milli\second}}

\newcommand{\licutliv}{\SI{48.7(2)}{\day}}

\newcommand{\lwcutmax}{\SI{1}{\second}}

\newcommand{\lwcutliv}{\SI{48.1(2)}{\day}}

\newcommand{\vpcutliv}{\SI{47.1(2)}{\day}}

\newcommand{\cvcutpemin}{\SI{10}{\pe}}
\newcommand{\cvcutenemin}{\SI{20}{\keV}}
\newcommand{\cvcutwin}{\SIrange[range-units = single]{-10}{200}{\ns}}

\newcommand{\cvcutacc}{\num{0.95}}   

\newcommand{\hvcutsliderwidth}{\SI{300}{\ns}}							
\newcommand{\hvcutdelpemax}{\SI{80}{\pe}}
\newcommand{\hvcutdelenemax}{\SI{150}{\keV}}
\newcommand{\hvcutdelwin}{\SI{8.8}{\us}}								

\newcommand{\hvcutlatepemax}{\SI{110}{\pe}}
\newcommand{\hvcutlateenemax}{\SI{200}{\keV}}

\newcommand{\hvcutlatectfpemax}{\SI{200}{\pe}}

\newcommand{\hvcutacc}{\num{0.94}}   
\newcommand{\npcutacc}{\mbox{$0.95^{+0.00}_{-0.01}$}}
\newcommand{\ttcutacc}{\mbox{$1.00^{+0.00}_{-0.01}$}}
\newcommand{\ttcuthalfwindow}{\SI{50}{\ns}}

\newcommand{\nscutacc}{\num{1.00}}   
\newcommand{\mfcutacc}{\num{0.99}}

\newcommand{\vwcutmax}{\num{0.20}}

\newcommand{\vwcutacc}{\num{1.00}}   

\newcommand{\uwcutmin}{\SI{100}{\pe}}

\newcommand{\uwcutacc}{\mbox{$0.99^{+0.01}_{-0.04}$}}%

\newcommand{\socutmin}{\SI{80}{\pe}}     
\newcommand{\socutenemin}{\SI{38}{\keV}}  

\newcommand{\socutmaxpaper}{\SI{460}{\pe}}
\newcommand{\socutenemaxpaper}{\SI{206}{\keV}}   

\newcommand{\dfcutmin}{\SI{40.0}{\micro\second}}
\newcommand{\dfcutmax}{\SI{334.5}{\micro\second}}

\newcommand{\dfcutzmin}{\SI{36.3}{\mm}}

\newcommand{\espeedbelowmesh}{\SI{0.93(1)}{\mm\per\us}}  

\newcommand{\dfcutfid}{\SI{36.9(6)}{\kg}}

\newcommand{\totalliv}{\vpcutliv}
\newcommand{\totalacc}{\mbox{$0.82^{+0.01}_{-0.04}$}}    

\newcommand{\dsfexpo}{\SI{1422(67)}{\kg\day}}

\newcommand{\dsfnumareventsinplot}{\num{1.5E7}}   

\begin{document}

\begin{frontmatter}







\title{First Results from the \dsf\ Dark Matter Experiment at Laboratori Nazionali del Gran Sasso}
\author[apc]{P.~Agnes}
\author[umass]{T.~Alexander}
\author[augustana]{A.~Alton}
\author[ucla]{K.~Arisaka}
\author[princeton]{H.O.~Back}
\author[fnal]{B.~Baldin}
\author[fnal]{K.~Biery}
\author[lngs]{G.~Bonfini}
\author[gssi]{M.~Bossa}
\author[mi]{A.~Brigatti}
% ...
\address[apc]{APC, Universit\'e Paris Diderot, Sorbonne Paris Cit\'e, Paris 75205, France}
\address[augustana]{Physics and Astronomy Department, Augustana College, Sioux Falls, SD 57197, USA}
% ...
\address[fnal]{Fermi National Accelerator Laboratory, Batavia, IL 60510, USA}
% ...
\address[gssi]{Gran Sasso Science Institute, L'Aquila 67100, Italy}
% ...
\address[lngs]{Laboratori Nazionali del Gran Sasso, Assergi (AQ) 67010, Italy}
% ...
\address[mi]{Physics Department, Universit\`a degli Studi and INFN, Milano 20133, Italy}
% ...
\address[princeton]{Department of Physics, Princeton University, Princeton, NJ 08544, USA}
% ...
\address[ucla]{Physics and Astronomy Department, University of California, Los Angeles, CA 90095, USA}
\address[umass]{Amherst Center for Fundamental Interactions and Physics Department, University of Massachusetts, Amherst, MA 01003, USA}
% ...
\fntext[kidnernote]{Deceased}
\cortext[cor1]{Corresponding author.}
\begin{abstract}
We report the first results of \dsf, a direct search for dark matter operating in the underground Laboratori Nazionali del Gran Sasso (LNGS) and searching for the rare nuclear recoils possibly induced by weakly interacting massive particles (WIMPs).  The dark matter detector is a Liquid Argon Time Projection Chamber with a \dsfactivemass\ active mass, operated inside a \lsvscintillatormass\ organic liquid scintillator neutron veto, which is in turn installed at the center of a \ctfwatermass\ water Cherenkov veto for the residual flux of cosmic rays.  We report here the null results of a dark matter search for a \dsfexpo\ exposure with an atmospheric argon fill.  This is the most sensitive dark matter search performed with an argon target, corresponding to a 90\% CL upper limit on the WIMP-nucleon spin-independent cross section of \dsflimit\ for a WIMP mass of \dsflimitmass.
\end{abstract}







\end{frontmatter}


\section{Introduction}
\label{sec:intro}

The matter content of the universe appears to be dominated not by ordinary baryonic matter but by a non-luminous and non-baryonic component: dark matter.  This surprising conclusion is derived from a wide range of observational evidence, ranging from studies of the internal motions of galaxies~\cite{Faber}, to the large scale inhomogeneities in the cosmic microwave background radiation~\cite{WMAP1}.  The precise nature of this matter is recognized as one of the most important questions in fundamental physics~\cite{napp}.

A favored candidate for the dark matter is a big-bang relic population of weakly interacting massive particles (WIMPs).  These could in principle be detected through their collisions with ordinary nuclei in an instrumented target, producing low-energy (\SI{<100}{\keV}) nuclear recoils~\cite{goodman}.  Very low interaction rates are expected for such particles, based on the model for their production and existing limits.  To detect these WIMPs, target masses of 0.1-10 tons may be required, and ultra-low background must be achieved by a combination of measures.  These include cosmic ray suppression by locating the experiments deep underground, selection of materials for low radioactivity, and instrumentation that can reject residual radioactive backgrounds in favor of the sought-after nuclear recoil events.

This paper reports the first physics data from the \dsf\ Liquid Argon Time Projection Chamber (\lar\ \tpc), operated in the Gran Sasso National Laboratory (LNGS) in Italy.  The \lar\ \tpc\ technique affords very strong background rejection by detecting both the scintillation light and the ionization electrons produced by recoiling nuclei \cite{warp,ds:ds-10-run3}.  \dsf\ is surrounded by a sophisticated water- and liquid scintillator-based veto system which further suppresses radiogenic and cosmogenic backgrounds.

The ultimate goal of \dsf\ is to conduct a background-free dark matter search with its 50-kg \tpc\ filled with argon derived from underground sources (\uar) \cite{ds:uar-extraction,ds:uar-distillation}, to reduce the rate of \ar\ decays in the active volume.  The present exposure amounts to \dsfexpo\ using an initial fill of atmospheric argon, obtained while the final purification of the UAr supply was still in progress. 
Atmospheric argon contains approximately \activityar\ of cosmogenic \ar~\cite{loosli,warp:39ar}.  Based on the measured upper limit of \ar\ in our \uar, a factor $>$\uardepletion\ below atmospheric argon~\cite{ds:uar-counting}, the present data contain \ar\ background equivalent to at least \dsfuareqexponoerr, or \dsfuareqexpoty, with \uar.  None of this background survives into the accepted event sample (see Sec.~\ref{sec:wimp_search}).  
This key result directly shows that the \ar\ background in the full \dsf\ run with \uar\ can be suppressed, and supports the claim that the order-of-magnitude larger \uar\ exposures envisioned for ton-scale \lar\ {\tpc s} can be free of \ar\ background.  

During this period, the liquid scintillator veto performance was limited due to the unexpectedly high content of \cfor\ in the trimethyl borate (\tmb) that was added as a neutron capture agent. 
Some of the \tmb\ feedstock was derived from modern carbon, which has a much higher \cfor\ content than petroleum-derived material.  The \tmb\ has since been removed and a source of low-activity \tmb\ identified.  The veto performance nevertheless has been adequate to measure and suppress the very low rate of neutron-induced events in the present data sample, another key goal of the \ds\ program.
\section{The \dsf\ Detectors}
\label{sec:detector}

\begin{figure}[t!]
\begin{center}
\includegraphics[height=0.7\columnwidth]{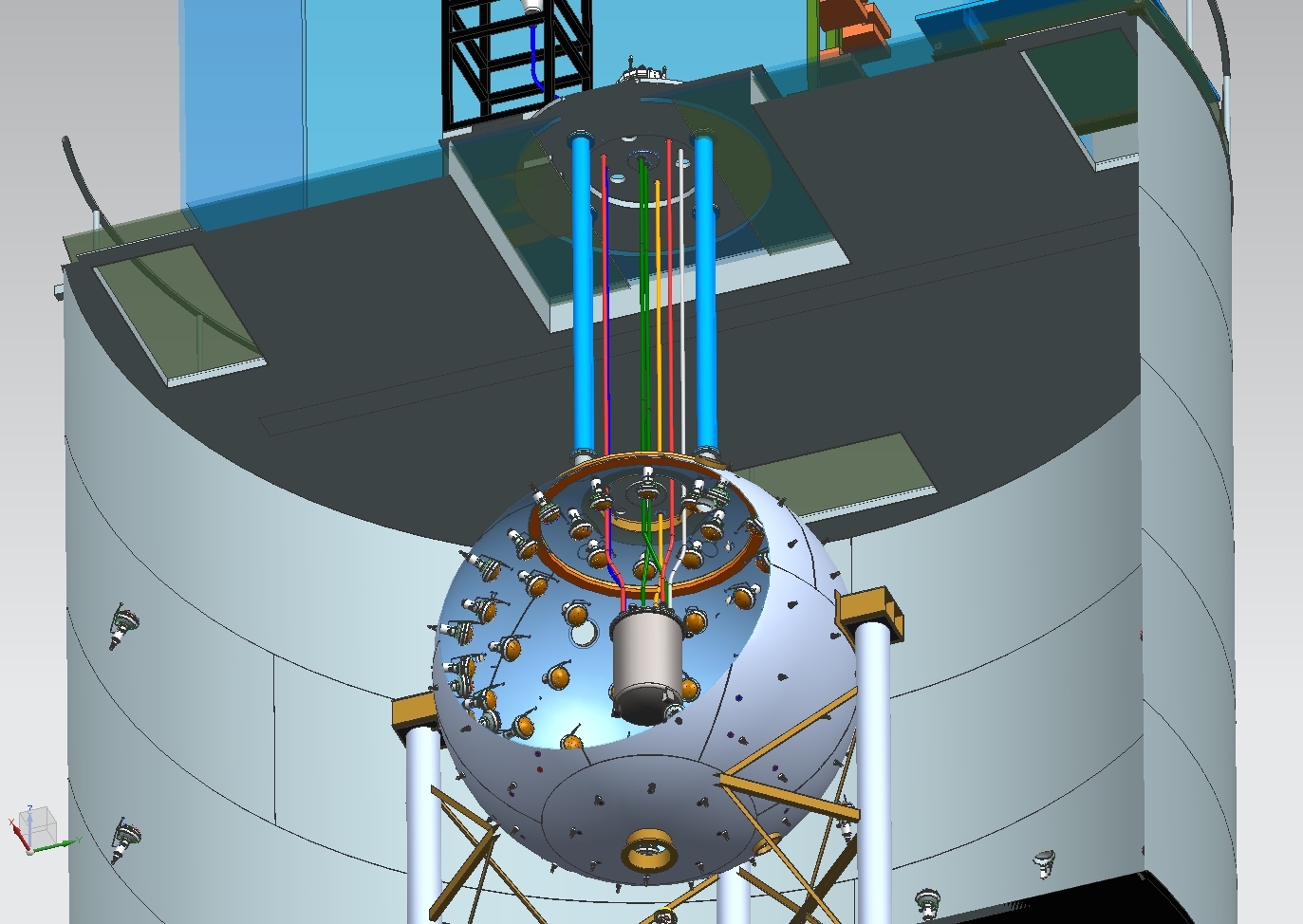}
\caption{The nested detector system of \dsf. The outermost gray cylinder is the \wcd, the sphere is the \lsv, and the gray cylinder at the center of the sphere is the \lar\ \tpc\ cryostat.}
\label{fig:ds-50-detectors}
\end{center}
\end{figure}

The \dsf\ apparatus consists of three nested detectors, see Fig.~\ref{fig:ds-50-detectors}.  From the center outward, the three detectors are: the Liquid Argon Time Projection Chamber, which is the dark matter detector; the Liquid Scintillator Veto (\lsv), serving as shielding and as anti-coincidence for radiogenic and cosmogenic neutrons, {\gr s}, and cosmic muons; and the Water Cherenkov Detector (\wcd), serving as shielding and as anti-coincidence for cosmic muons~\cite{ctf:results,ctf:scitech}.  The detector system is located in Hall C of LNGS at a depth of \lngsequivdepth\ \cite{lngs-depth}, in close proximity to and sharing many facilities with, the Borexino solar neutrino detector~\cite{bx:detector,bx:plants}.  

The \lar\ \tpc\ can exploit pulse shape discrimination and the ratio of scintillation to ionization to reject \bg\ background in favor of the nuclear recoil events expected from WIMP scattering~\cite{boulay,warp}.  It can also exploit the \tpc's spatial resolution to reject surface backgrounds and to reject multi-sited events.  Events due to neutrons from cosmogenic sources and from radioactive contamination in the detector components, which also produce nuclear recoils, are suppressed by the combined action of the neutron and cosmic ray vetoes.  The liquid scintillator also provides additional rejection of $\gamma$-ray background from the detector materials.  The water-plus-liquid scintillator design was motivated in part by the success of this shielding concept in achieving very low backgrounds in Borexino~\cite{bx:detector,bx:7be-precision,bx:phase-I}.

The \wcd\ is an \ctfdiameter-diameter, \ctfheight-high cylindrical tank filled with high purity water.  The tank was originally part of the Borexino Counting Test Facility.  The inside surface of the tank is covered with a laminated Tyvek-polyethylene-Tyvek reflector~\cite{daya-bay}.  An array of \ctfpmtnum\ \ctfpmt\ \ctfpmtsize~{\pmt s}, with \ctfpmtqe\ average quantum efficiency (QE) at \ctfpmtwave, is mounted on the side and bottom of the water tank to detect Cherenkov photons produced by muons or other relativistic particles traversing the water.

The \lsv\ is a \lsvdiameter-diameter stainless steel sphere filled with \lsvscintillatormass\ of borated liquid scintillator.  The scintillator consists of equal amounts of pseudocumene (PC) and trimethyl borate (TMB), with the wavelength shifter Diphenyloxazole (PPO) at a concentration of \lsvppoconcentration.  The sphere is lined with Lumirror~\cite{lumirror} reflecting foils.  An array of  \lsvpmtnum\ \lsvpmt\ \lsvpmtsize~{\pmt s}, with low-radioactivity glass bulbs and high-quantum-efficiency photocathodes (\lsvpmtqe\ average QE at \lsvpmtwave), is mounted on the inside surface of the sphere to detect scintillation photons.

The neutron-capture reaction \borten($n$,$\alpha$)\lith\ makes the borated scintillator a very effective veto of neutron background~\cite{wright}.   The \tmb, \tmbchem, contains $^{\rm nat}$B which has a \abborten\ natural abundance of \borten\ with its large (\lsvbtenxsec) thermal neutron capture cross section.  The thermal neutron capture time in the borated scintillator is calculated to be just \lsvtmbfiftypercentcapturetime, compared to \lsvtmbzeropercentcapturetime\ for pure \pc~\cite{bx:detector}.

The \borten\ neutron capture proceeds to the \lith\ ground state with branching ratio \brbortenground, producing a \enbortengroundalpha\ $\alpha$ particle, and to a \lith\ excited state with branching ratio \brbortenexcited\, producing a \enbortenexcitedalpha\ $\alpha$ particle and a gamma-ray of \enbortenexcitedgamma.  Because of quenching, the scintillation light output of the capture to \lith(g.s.) is expected to be in the \bg-equivalent range \lsvalphaequivenergy~\cite{greenwood,wang}.  
Preliminary measurements with our scintillator appear consistent with this expectation.
The measured \lsv\ photoelectron (\si{\pe}) yield is \lsvly, making this quenched energy readily detectable.  The high \cfor\ decay rate in the \lsv\ and the fact that its spectrum covers the signal expected from the $\alpha$'s from neutron capture on \borten\ severely reduced the effectiveness of the neutron veto in the present data set.  The rejection power is estimated from simulations to be \odneutronrejectionachieved\ instead of the design value of~\odneutronrejectiondesign~\cite{wright}.

\begin{figure}[t!]
\begin{center}
\includegraphics[height=4.0in]{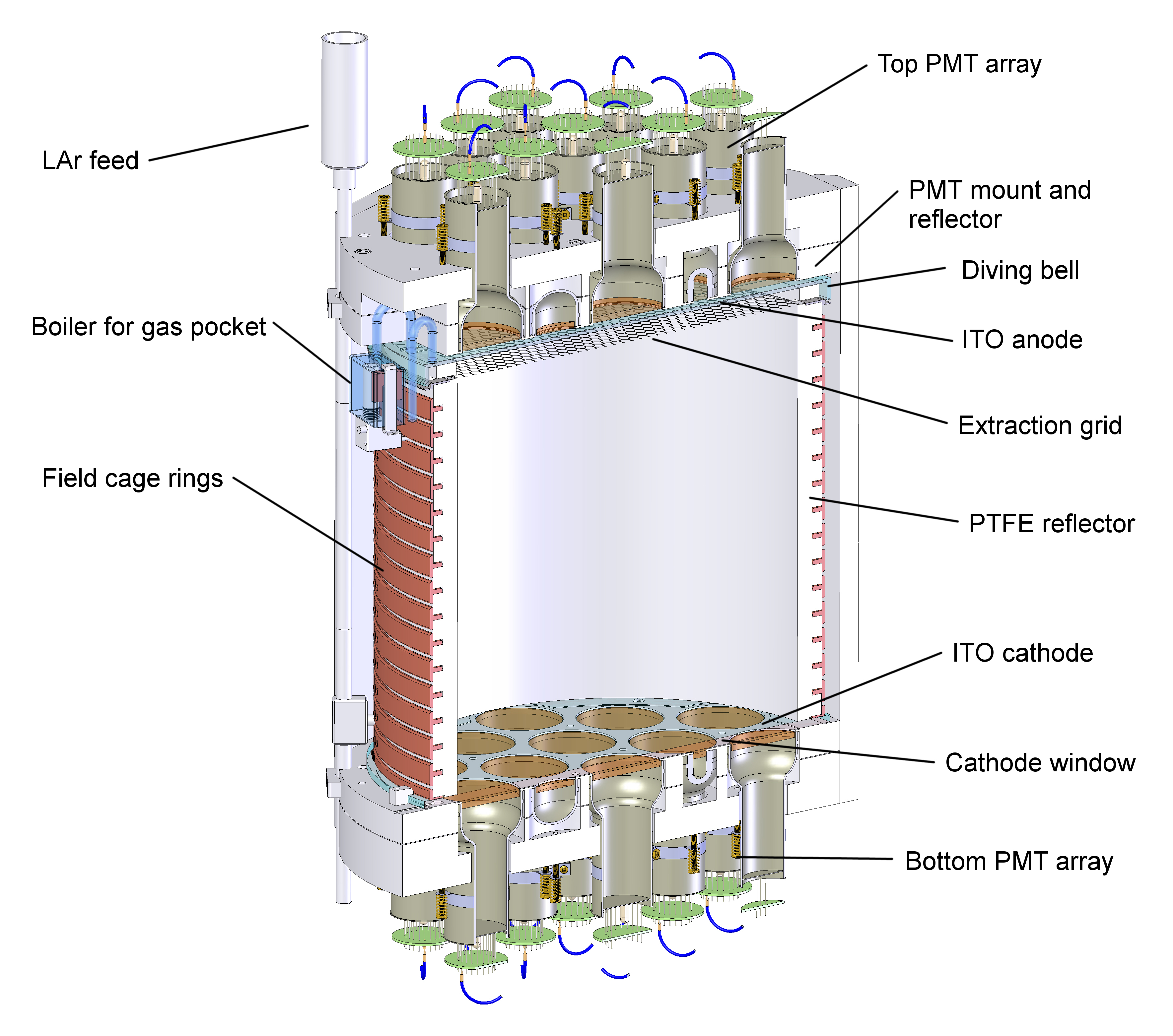}
\caption{The \dsf\ Liquid Argon Time Projection Chamber.}
\label{fig:ds-50-tpc}
\end{center}
\end{figure}

The \dsf\ \tpc, as shown in Fig.~\ref{fig:ds-50-detectors}, is contained in a stainless steel cryostat that is supported at  the center of the \lsv\ on a system of leveling rods.  Its design was based on that of the \dst\ prototype, which operated for \dstexp\ at LNGS~\cite{ds:ds-10-run3}. 
A cut-away view of of the \tpc\ is given in Fig.~\ref{fig:ds-50-tpc}.

Ionizing events in the active volume of the \lar\ \tpc\ result in a prompt scintillation signal called ``\sone''.  Ionization electrons escaping recombination drift in the \tpc\ electric field to the surface of the \lar, where a stronger electric field extracts them into an argon gas layer between the \lar\ surface and the \tpc\ anode.  The electric field in the gas is large enough to accelerate the electrons so that they excite the argon, resulting in a secondary scintillation signal, ``\stwo'', proportional to the collected ionization.  Both the scintillation signal \sone\ and the ionization signal \stwo\ are measured by the same \pmt\ array.  The temporal pulse shape of the \sone\ signal provides discrimination between nuclear-recoil and electron-recoil events.  The \stwo\ signal allows the three-dimensional position of the energy deposition to be determined and, in combination with \sone, provides further discrimination of signal from background.   A significant fraction of events also exhibit an ``\sthree" signal.  The S3 pulse resembles \stwo\ in pulse shape but is typically \dsfstwooversthree\ times smaller and always follows \stwo\ by a fixed delay equal to the maximum drift time in the \lar\ \tpc.  \sthree\ is believed to result from electrons released from the cathode (at the bottom of the TPC) when struck by the bright \stwo\ UV light.

The active \lar\ is contained in a cylindrical region viewed by \dsfpmtnum\ \dsfpmt\ \dsfpmtsize\ low-background, high-quantum-efficiency  {\pmt s}, nineteen each on the top and the bottom.  The average quantum efficiency of the {\pmt s} at room temperature is \dsfpmtqe\ at \dsfpmtwave.  The  {\pmt s} are submerged in liquid argon and view the active \lar\ through fused-silica windows, which are coated on both faces with transparent conductive indium tin oxide (\ito) films \dsfitothickness\ thick.  This allows the inner window faces to serve as the grounded anode (top) and $-$HV cathode (bottom) of the \tpc\ while maintaining their outer faces at the average \pmt\ photocathode potential.  The cylindrical wall is a \dsfthickptfereplector-thick \ptfe\ reflector fabricated with a modified annealing cycle  to increase its reflectivity.  The reflector and the windows at the top and bottom of the cylinder are coated with a wavelength shifter, tetraphenyl butadiene (\tpb), that absorbs the \arwave\ scintillation photons emitted by liquid argon and re-emits visible photons (peak wavelength \dsfpmtwave) that are reflected, transmitted, and detected with high efficiency.  The thickness of the \tpb\ coating on the windows varies from \dsftpbcenterthickness\ at the center to \dsftpbedgethickness\ at the edge of the active volume.  The thickness of the \tpb\ on the cylindrical wall is \dsftpbwallhalfthickness\ at half-height and \dsftpbwallbottomthickness\ at the top and bottom.

The fused silica anode window has a cylindrical rim extending downward to form the ``diving bell'' that holds the \dsfgaspocketthickness-thick gas layer of the \tpc, produced by boiling argon within the cryostat (outside the \tpc\ active volume) and delivering the gas to the diving bell.  The gas then exits the bell via a bubbler that maintains the \lar/gas interface at the desired height.

The electron drift system consists of the \ito\ cathode and anode planes, a field cage, and a grid that separates the drift and electron extraction regions.  The grid, \dsflarovergrid\ below the liquid surface, is a hexagonal mesh etched from a \dsfgridfoil-thick stainless steel foil and has an optical transparency of \dsfgridfoiltrasparency\ at normal incidence.  
Voltage is applied between the cathode and grid to produce a vertical electric field to drift the ionization electrons upward.  Outside the cylindrical \ptfe\ wall, copper rings at graded potentials keep the drift field uniform throughout the active volume.  An independently-adjustable potential between the grid and anode creates the fields that extract the electrons into the gas and accelerate them to create the secondary scintillation signal.  The data reported here were taken with a \dsfcathodepot\ cathode potential and a \dsfgridpot\ grid potential, giving drift, extraction, and electroluminescence electric fields of \dsfdriftfield, \dsfextractionfield, and \dsfmultfield, respectively.  The choice of drift field was dictated by the results of the calibration experiment \scene, which uncovered a drift-field-induced quenching of the S1 light yield for nuclear recoils~\cite{scene1,scene2}.  The maximum drift time is \dsftdriftmaxanal, and the measured value of the drift speed is \espeedbelowmesh.

The active \lar\ volume is bounded by the cylindrical \ptfe\ wall, the cathode, and the grid.  When warm, it is \dsfactivevolumediameter\ in diameter and \dsfactivevolumeheight\ in height.  This gives an active mass when cold of \dsfactivemass\ of liquid argon, where the uncertainty is primarily in the thermal contraction of the \ptfe.

Cooling of the cryostat is done using an external circulation loop.  Argon gas drawn from the cryostat at \dsfcryoflowrate\ passes out of the detector system to the cryogenic and purification system, located in the radon-suppressed clean room, which contains all equipment interfacing directly to the detectors.  The gas passes through a SAES Monotorr PS4-MT50-R-2 getter~\cite{saes}, which reduces contaminants such as \ot\ and \nt\ to sub-ppb levels.  The gas is then pre-cooled in a heat exchanger before passing through a cold-charcoal radon trap that is operated in the range \dsfcryorntrapartemp.  The argon is then liquefied by a liquid-nitrogen-cooled heat exchanger.  The loop cooling power is controlled to maintain a stable pressure in the cryostat.  The pressure oscillates within a band of \dsfcryopressurestability\ around the set point of \dsfcryopressure.  The electron mean drift lifetime as measured through \stwo/\sone\ vs.~drift time was \dsfelectronmeanlifefirst\ for the initial set of runs acquired in October-November 2013.  For the runs acquired in~2014, which provide the large majority of the exposure, the electron mean life was \dsfelectronmeanlifesecond.

The cryogenic and purification system includes a \krthree\ source~\cite{venos}, whose use in \lar\  detectors was introduced in Ref.~\cite{lippincott-kr}. 
The argon flow can be directed through the source to introduce \krthree\ into the TPC for calibration of the energy response of the detector.  \krthree\ is produced by the decay of \rbthree\ ($\tau$=\rbthreetau), which was prepared in the form of RbCl and adsorbed on a pellet of synthetic activated charcoal.  
The activity of \rbthree\ when the source was prepared (September 2012) was \SI{8.5}{\kilo\becquerel}.
While \rbthree\ is firmly adsorbed onto the activated charcoal, \krthree\ escapes into the recirculation stream, and, after passing through a \SI{0.5}{\um} filter and the radon trap, flows to the TPC. 
The \krthree\ decays with $\tau$=\krthreetau\ to the ground state in two sequential electromagnetic transitions of \krthreefirstene\ and \krthreesecondene\ energy with an intermediate mean life of about \krthreeinttau.  Because of the slow component of \lar\ scintillation, the \tpc\ is unable to resolve the two decays and sees a single deposition of \krthreepeakene.  

In liquid argon, scintillation is initiated both by excitation and by recombination after ionization.  The \arwave\ scintillation photons are emitted from two nearly degenerate excimer states, a long-lived (\longlivedtriplestatedimerargon) triplet state, and a short-lived (\shortlivedtriplestatedimerargon) singlet state.  The difference in ionization density between nuclear recoils (from WIMP or neutron scattering) and electron recoils (from \bg\ radiation) produces a significant difference in the radiative decay ratio of these states and hence in the time profile of the \sone\ scintillation light~\cite{kubota,hitachi}.  Nuclear recoils have more of the fast scintillation component than electron recoils, providing a very powerful ``pulse shape discrimination'' (PSD) between electron backgrounds and nuclear-recoil signals~\cite{boulay}.  In the analysis presented here we use a simple PSD parameter, \fno, defined as the fraction of the \sone\ signal (defined hereafter as the integral of the \sone\ pulse over \fixedintone, see Sec.~\ref{sec:recon}) that occurs in the first \timefno\ of the pulse, which is typically~$\sim$\num{0.3} for \bg-events and~$\sim$\num{0.7} for nuclear recoils.  For \bg-events, the low density of electron-ion pairs also results in less recombination and therefore more free electrons, compared to a nuclear recoil track of the same \sone~\cite{kubota,doke,xenon93}.  The ratio of ionization (measured by S2) to scintillation (S1) can therefore also be used to distinguish electron recoils from nuclear recoils. In this paper, we use PSD and basic cuts on \stwo\ to reduce backgrounds, but we do not yet exploit the discrimination power of \stwo/\sone.

\section{Electronics and Data Acquisition}
\label{sec:daq}

The electronics and data acquisition (DAQ) are divided into two main sub-systems, one for the \tpc\ and one for the vetoes~\cite{ds:daq}.  In the \tpc\ sub-system, the PMTs use resistive divider circuits on Cirlex substrates. The anode signal from each of the \tpc\  {\pmt s} is first amplified by a cryogenic preamplifier on the \pmt\ voltage divider, immersed in liquid argon.  This allows the  {\pmt s} to be operated at low gain (typically \dsfpmtgain), reducing the occurrence of the sporadic light emission we have observed in R11065s, while maintaining a very high signal-to-noise ratio.  
The fast pulse pre-amplifier (\dsfcoldampbandwith\ bandwidth with extended low frequency response down to \dsfcoldamplowfreqtimeconst) has an internal gain of \dsfcoldampgain\ and is coupled to the anode of the PMT with a 200~$\Omega$ load. This value optimizes the impedance matching, giving an extra gain of 8 V/V with respect to the 25~$\Omega$ load that is normally used with a passive PMT readout (\dsfcoldamptermination\ on the anode and \dsfcoldamptermination\ input to the amplifier). The output swing of the back-terminated output is \dsfcoldamppeaktopeak, corresponding to about 1500 \si{\pe}, with a noise equivalent of \dsfcoldampnoise\ (referred to the output). At room temperature, the signal is then further amplified $\times$\dsffemgain\ and split, with one copy sent to a high speed discriminator, set to \dsftriggereneth\ and used to form the \tpc\ trigger.  A second copy is filtered and sent to a  \dsfdigitresolution, \dsfdigitsamplespeed\ digitizer channel (CAEN~1720~\cite{caen}).

In the veto subsystem, the anode signals from the \odpmtnum\ {\pmt s} undergo amplification and splitting by means of a custom front-end board.  A $\times$\odfemgain\ amplified signal is sent to \odpmtnum\ channels of NI PXIe-5162 digitizer~\cite{ni} which sample at \oddigitsamplespeed\ with a \oddigitresolution\ resolution.  Zero-suppression is performed on the fly and only sections of the waveform around identified peaks above threshold are stored.  The zero-suppression threshold was set to a level about \odzerosuppressionrun\ times the amplitude of a single-photoelectron pulse for routine data taking.  

The data readout in the three detector subsystems is managed by dedicated trigger boards: each subsystem is equipped with a user-customizable FPGA unit (CAEN V1495~\cite{caen}), in which the trigger logic is implemented.  The inputs and outputs from the different trigger modules are processed by a set of electrical-to-optical converters (Highland V720 and V730~\cite{highland}) and the communication between the subsystems uses dedicated optical links.  
To keep the TPC and the Veto readouts aligned, a pulse per second (PPS) generated by a GPS receiver is sent to the two systems, where it is acquired and interpolated with a resolution of \gpsabsoluteaccuracy\ to allow offline confirmation of event matching.

The DAQ sub-systems are handled by a common run controller that can be configured to permit different acquisition modes -- either sharing a global trigger among all three detectors, or allowing independent triggers.  In the shared global-trigger configuration, used for the physics data set, the photomultiplier signals from the \tpc\ and veto are stored when at least three \tpc\  \pmt\ discriminators give signals within a \dsftriggerwin\ window.  
This trigger has an efficiency \dsftrigeff\ for \sone$>$\dsftrigeffthresh.

After some initial running, the trigger was modified to reject the high-energy part of the \ar\ energy spectrum to save disk space.  
To do this, the FPGA calculating the trigger condition counts the number of discriminator firings in the \tpc\  in the \SI{5}{\micro\second} after the lower-level trigger and issues a flag for events above $\sim$\SI{600}{\pe}, outside of the WIMP region of interest.
The flagged events are pre-scaled by a factor \num{33}, reducing the trigger rate from the \dsftrigrate\ of the low-level trigger to 
 \dsftrigrategtwo. 

The TPC data acquisition window was \dsfacquiwindow\ for routine data taking, and, to reduce re-triggers on the tail of \stwo\ or on \sthree, we employed an \dsfinhibitwindow\ inhibit after each trigger, during which the DAQ would not accept a new trigger.
Although a mean neutron capture time of \lsvtmbfiftypercentcapturetime\ is expected in the boron-loaded liquid scintillator, Monte Carlo simulations show that some neutrons that interact in the \tpc\ will capture in surrounding materials as long as several tens of \si{\micro\second} after the interaction in the \tpc~\cite{wright}.  Therefore, veto acquisition windows as long as \odacquisitionwin\ are used in order to include possible delayed neutron captures.
\section{\pmt\ Calibration}
\label{sec:pmt_calib}

The measurement of the \pmt\ gains and the study of the PMT charge response are performed for the three detector subsystems by injecting light from pulsed laser diodes into their respective sensitive volumes through optical fibers.  Optical filters are used to attenuate the laser intensity to provide an average occupancy of $<$\dsflaserpmtoccupancy\ on most {\pmt s}.  
 
In the \tpc, laser pulses have $\sim$\dsflaserpulsewidth\ duration at \dsflaserwavelength\ wavelength. 
Simultaneous triggers are sent at \dsflaserpulserate\ to the laser and the DAQ, which uses a \dsflaseracquiwindow\ acquisition window for these laser runs. 

We estimate the mean and variance of the single photoelectron (SPE) response for each channel using a statistical method that does not make any assumptions about the shape of the SPE charge spectrum. Two separate charge spectra are constructed, one from a \dsflaserintegrationwindow\ wide signal window around the arrival time of the laser pulses and a second from a pedestal window of the same width, but offset in time -- where no signal is expected. The mean of the charge spectrum in the signal window, \laserspectrummean, can be expressed as \laserspectrummean$=$\laserpedmean$+$\laserspemean$\cdot$\laserpemean, where \laserpedmean\ is the mean of contributions unrelated to the signal, including electronics noise; \laserspemean\ is the mean of the SPE charge distribution; and \laserpemean\ is mean number of photoelectrons produced per laser pulse. In order to estimate \laserspemean, \laserspectrummean\ is obtained directly from the charge spectrum of the signal window, \laserpedmean\ from the pedestal window, and \laserpemean\ using a combination of the two spectra, assuming the number of photoelectrons follow a Poisson distribution. We use the same statistical procedure to estimate the variance. Using Monte Carlo simulations of fake single photoelectron signals overlaid on true electronics baselines, we estimate that the systematic uncertainty of this method in determining the SPE mean is $<$2.5\%.
 
\tpc\ laser runs are taken at least daily and, over the course of \totalcalendartime, the \si{\spe} mean decreased by \dsflaserspemeanstability\ (\pmt\ bias voltages were unchanged).  The \tpc\ light yield (\si{\pe/\keV}) measured from the endpoint of the \ar\ $\beta$ spectrum was stable to within \dsflystability\ over the period of all data taking.

Calibration of the \si{\spe} charge response for \lsv\ and \wcd\ {\pmt s} follows a similar procedure. An optical fiber in front of each \pmt\ injects low intensity laser light at \odlaserpulserate\ with simultaneous triggers to the laser and veto DAQ, and a charge spectrum is constructed by integrating the laser pulse found by the zero-suppression algorithm over a time window of 150~ns around the laser trigger.
 A single Gaussian is fit to the \si{\spe} peak of the charge spectrum for each channel, and the fitted Gaussian mean is taken to be the mean \si{\spe} response. 

\section{TPC Event Reconstruction}
\label{sec:recon}

The TPC event reconstruction software is built within the Fermi National Accelerator Laboratory's {\it art} framework~\cite{art}. 
During normal data taking, raw waveforms from the \tpc\ and vetoes are separately analyzed to reconstruct the physical pulses in each detector. For each channel of the \tpc, a baseline is determined and subtracted from the raw waveform. To account for slow variations, the baseline is defined as a moving average, with window length \movingaveragewindow, in regions of the waveform consistent with only electronic noise. In regions with sharp excursions (such as single photoelectrons or scintillation pulses), the baseline is linearly interpolated between the two nearest quiet regions. The moving baseline algorithm is effective at reducing noise contributions to integrated signal estimations, which is important for PSD.

The baseline-subtracted waveforms of each channel are then scaled by the corresponding \si{\spe} mean, zero-suppressed with a threshold of \sumchzerosuppthresh/sample, and added together to form a sum channel, which is used for pulse finding. The use of zero-suppression is intended to reduce the effects of coherent noise across all channels. The pulse finding algorithm is general and can find both \sone\ and \stwo\ with high efficiency. The algorithm does a coarse-grained search to discern pulses and a fine-grained search to identify pulse start times, using a threshold of \pulsestartthresh/sample. It is also adept at distinguishing overlapping pulses such as multiple \stwo\ signals from multi-sited depositions.

Using the start time found for each pulse, two integrals, with integration lengths \fixedintone\ and \fixedinttwo, are computed on the scaled baseline-subtracted waveform of each channel (without zero-suppression). The integrals are then summed across all channels to produce the total number of photoelectrons observed in each pulse. The \fixedintone\ integration window is used for \sone\ pulses and the \fixedinttwo\ length for \stwo\ pulses. The use of fixed length integration windows simplifies electronic noise considerations, especially for \fno\ distributions. 

For events with two pulses found, the earlier pulse is assumed to be \sone\ and the later to be \stwo.  Subsequent cuts establish the validity of these assumptions event-by-event.  For events with three pulses found and the time difference between the second and third pulse approximately equal to the maximum drift time of the \tpc, the first pulse is assumed to be \sone, the second to be \stwo, and the third to be \sthree.

Several corrections are applied to the \sone\ and \stwo\ integrals to account for geometrical variations of light production and collection. 

Due to total internal reflection at the liquid surface, light collection of scintillation pulses varies by \sonezvariation\ between the top and bottom of the \tpc.  An empirical $z$-dependent correction, derived from \krthree\ and \ar\ calibration data and normalized to the center of the \tpc, is applied to \sone.  

Electronegative impurities in the \lar\ capture drifting electrons.  
This results in the number of drifting electrons, and hence \stwo, decreasing exponentially with the time taken to drift between the interaction point and the liquid-gas interface.
We fit for this electron drift lifetime, then correct \stwo, normalizing to the top of the \tpc.  Our very high electron mean drift lifetime induces a maximum \stwozvariation\ correction for the runs acquired through February~2014.  During a gap in the data taking that followed, the lifetime continued to improve.  We do not apply any correction to data collected after this period, about 75\% of the total, since the electron drift lifetime has become too long to be measured reliably.

We discovered during commissioning that the amplitude of \stwo\ has a strong radial dependence, where events under the central \pmt\ exhibit greater than three times more electroluminescence light than events at the maximum radius.  
Only basic cuts on S2 are used in the present analysis, and this does not affect our results.
Preliminary $x$-$y$ reconstruction algorithms indicate that the radial variations can be empirically corrected using calibration data.

\section{Veto Event Reconstruction}
\label{sec:vetorecon}

Due to the use of DAQ-level zero-suppression, reconstruction of \lsv\ and \wcd\ signals is different from the \tpc\ reconstruction. Pulses are naturally defined as the non-zero portion of each raw waveform for each channel. 
The DAQ records \odpresamps\ (\odpresampstime) before and after each pulse. The first \odbaselinewindow\
before the pulse are averaged to define a baseline, which is subtracted from 
the waveform.
Each channel is then scaled by the corresponding \si{\spe} mean and the channels in each veto detector are summed together. 

A clustering algorithm on the sum waveform identifies physical events in the \lsv.  To handle the high pile-up rate due to \cfor, the algorithm is a ``top-down'' iterative process of searching for clusters from largest to smallest. These clusters are used only for building the \cfor\ and \coba\ spectra and determining the light yield of the \lsv. Identification of coincident signals between \lsv\ and \tpc\ uses fixed regions of interest of the sum waveform and is described in Sec.~\ref{sec:analysis}. For tagging of muons in the \lsv\ and \wcd, the total integrated charge of each detector is used.

\begin{figure}[t!]
\begin{center}
\includegraphics[width=\columnwidth]{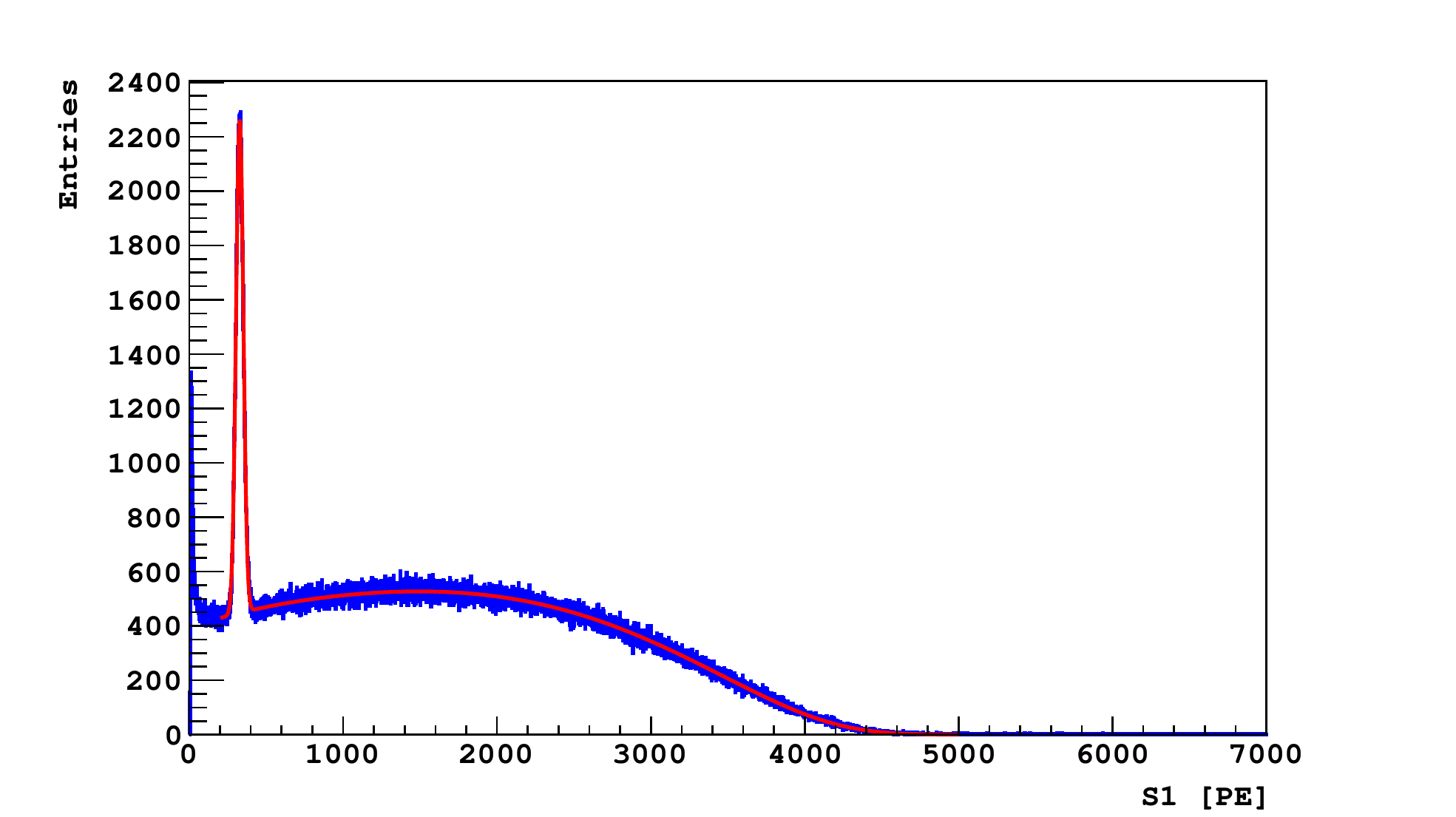}
\caption{The primary scintillation (\sone) spectrum from a zero-field run of the \dsf\ \tpc.  Blue: \sone\ spectrum obtained while the recirculating argon was spiked with  \krthree, which decays with near-coincident conversion electrons summing to \krthreepeakene. Red: fit to the \krthree+\ar\ spectrum, giving a light yield of \dsfnullfieldly\ at zero drift field.
}
\label{fig:dsfdriftfieldly} 
\end{center}
\end{figure}

\section{TPC Energy Calibration and Light Yield }
\label{sec:energy}

In the data presented here, taken with atmospheric argon, the TPC trigger rate is dominated by \ar\ $\beta$ decays, with their \epar\ endpoint.  The spectrum observed in the presence of the \krthree\ source at zero drift field, clearly dominated by \ar\ decay, is shown in Fig.~\ref{fig:dsfdriftfieldly}.
The measured rate of \krthree\ events is \dsfkrthreeobsrate.
Because it affects the optics of the detector, the gas pocket was maintained even when operating the detector at zero drift field to collect reference data for light yield.  
The spectrum is fit to obtain the measurement of the light yield of the detector at the \krthreepeakene\ reference line of \krthree.  
The fit of the entire spectrum, encompassing the  \ar\ and  \krthree\ contributions, shown in Fig.~\ref{fig:dsfdriftfieldly}, returns a light yield of \dsfnullfieldly\ at zero drift field, after including systematic errors. Fitting the light yield from the \krthree\ peak alone gives the same result within the fitting uncertainty.
The resolution is about 7\% at the \krthree\ peak energy.
The dominant uncertainty in the \krthree\ light yield comes from the systematic uncertainties on the mean SPE response of the PMTs. 
Table \ref{tab:ly} summarizes the values for the light yield from \krthree\ with and without drift field. 
Note that the value of the light yield has a larger uncertainty at zero field, where it is neither possible to account for non-uniformities in the \krthree\ distribution in the active volume nor to correct for the $z$-dependent light collection variations described in Sec.~\ref{sec:recon}. Both effects are accounted for with the drift field on, where $z$-position information is available.

\begin{table*}[t!]
\begin{tabular*}{\textwidth}{@{\extracolsep{\fill}}ccccc}
\hline\hline
Experiment			&Drift Field	&Light Yield	\\ 
\hline
\dsf\					&\dsfdriftfield	&\dsfdriftfieldly	\\  
\dsf\					&Zero		&\dsfnullfieldly	\\  
\hline
\scene\ (Jun~2013 Run)	&Zero	&\scenenullfieldlyjun	\\  
\scene\ (Oct~2013 Run)	&Zero	&\scenenullfieldlyoct	\\  
\hline\hline
\end{tabular*}
\caption{Reference \krthree\ values of the light yields from \scene\ and \dsf\ used to correlate the \sone\ and \stwo\ scales of \dsf\ with the \scene\ calibration.
Note that most of the systematic errors are correlated between the \dsf\ \dsfdriftfield\ and zero-field light yields.}
\label{tab:ly}
\end{table*}

\begin{table*}[t!]
\begin{center}
\begin{tabular*}{0.8\textwidth}{@{\extracolsep{\fill}}cccccc}
\hline\hline
			Energy [\si{keV}]
						&$\mathcal{L}_{\rm eff,\,\krthree}$
										&\sone$_{\rm DS-50}$ [\si{\pe}]
\\
\hline
			\num{16.9}	&\num{0.202(8)}	&\num{27.0(18)}	\\ 
			\num{20.5}	&\num{0.227(10)}	&\num{36.8(27)}	\\  
			\num{25.4}	&\num{0.224(10)}	&\num{45.0(33)}	\\  
			\num{36.1}	&\num{0.265(10)}	&\num{75.7(50)}	\\  
			\num{57.2}	&\num{0.282(13)}	&\num{127.6(91)} \\  
\hline\hline
\end{tabular*}
\end{center}
\caption{Expected nuclear recoil responses of \dsf\ based on \scene\ calibration.  Second column: $\mathcal{L}_{\rm eff,\,\krthree}$ is the quenching of nuclear recoils in \lar\ at \dsfdriftfield, relative to the yield of \krthree\ at zero field, as measured and defined in Ref.~\cite{scene2}.  Values are reported for all energy points of nuclear recoils examined in \scene~\cite{scene2}.  Third column: \sone\ yields of nuclear recoils measured in SCENE at \dsfdriftfield, and projected to \dsf\ by using the cross-calibration response of \krthree.  
}
\label{tab:scene}
\end{table*}

To obtain the best calibration of the response in \sone\ and \stwo\ for nuclear recoils needed for the \ds\ program, members of the collaboration and others performed an experiment called \scene~\cite{scene1,scene2}.  The \scene\ experiment measured the intrinsic scintillation and ionization yield of recoiling nuclei in liquid argon as a function of applied electric field by exposing a small \lar\ \tpc\ to a low energy pulsed narrowband neutron beam produced at the Notre Dame Institute for Structure and Nuclear Astrophysics.  Liquid scintillation counters were arranged to detect and identify neutrons scattered in the \tpc, determining the neutron scattering angles and thus the energies of the recoiling nuclei.  The use of a low-energy narrowband beam and of a very small \tpc\ allowed \scene\ to measure the intrinsic yields for single-sited nuclear recoils of known energy, which is not possible in \dsf.

The measurements performed in \scene\ were referenced to the light yield measured with a \krthree\ source at zero field.  The use of the same \krthree\ in \dsf\ allows us to use the relative light yields of the two experiments (see Table \ref{tab:ly}) to determine, from the \scene\ results, the expected \sone\ and \stwo\ signals of nuclear recoils in \dsf.
Table~\ref{tab:scene} summarizes the expected \dsf\ \sone\ yields derived with this method and thus provides nuclear recoil energy vs.\ \sone\ for \dsf.
For the analysis reported here, we interpolate linearly between the measured \scene\ energies and assume that $\mathcal{L}_{\rm eff,\,\krthree}$ is constant above a nuclear recoil energy of \SI{57.2}{\keV}.

As part of the \dsf\ program, we planned to deploy both gamma and neutron sources in the \lsv\ near the cryostat for calibrations over a broad range of energies and for direct measurements of the \tpc\ response to nuclear recoils and of the \lsv\ response to neutrons.  The equipment needed to deploy such sources through the \wcd\ and \lsv\ was not available during the running and analysis reported here.  It has been recently completed and commissioned, and analyses for all these purposes are underway.

\section{LSV Energy Calibration and Light Yield }
\label{sec:lsvly}

The neutron veto light yield is measured by fitting the \cfor\ and \coba\ spectra.  The high rate in the \lsv, from \cfor\ and a low-energy, low-PMT-multiplicity signal that decreased sharply during the run, requires clustering and pulse selection before fitting these spectra.  
For the \cfor\ fit, the fit parameters are the \cfor\ rate, the light yield, the pedestal mean and variance, and a constant term.  We use the variance of the \si{\spe} charge spectrum from the \si{\spe} calibration as a fixed parameter.  A light-yield variance term that includes geometrical variations is determined empirically and kept fixed in the fit.  The high rate of \cfor\ decays, \lsvcforrate, requires allowance for pileup in the fit function. 
We also studied the light yield at higher energies by observing coincident 
\gr\ events in the TPC and the LSV from \coba\ decays in the cryostat steel, 
which dominate the rate above 1~MeV. The fit function was a single Gaussian (the two gammas are not resolved) with an exponential component modeling the  background in the coincidence window. 

By combining the \cfor\ and \coba\ results, we obtain an LSV light yield of \lsvly. 
The error quoted on the light yield includes systematic uncertainties from the cuts used to suppress the low-energy background, the clustering algorithm, and the observed variation during the run.
In the current analysis, the \lsv\ light yield is used only to infer the threshold of the veto cuts described in Sec.~\ref{sec:analysis} to estimate the neutron background rejection via Monte Carlo studies.

\section{Data Analysis}
\label{sec:analysis}

The goal of the analysis is to distinguish events that are induced by the scattering of WIMPs in the active \lar\ from those caused by any other process.  The signature of a WIMP scattering event is a single-sited nuclear recoil (NR), that is, an energy deposition in one location in the TPC with observed properties consistent with a heavy recoiling particle (the argon atom) and no activity in the vetoes.
The dominant backgrounds by far are those from \bg\ decays in the materials of the TPC and cryostat.  In the atmospheric argon used for the analysis reported here, the overwhelming majority of these events are due to $\beta$~decay of \ar, as can be seen in Fig.~\ref{fig:dsfdriftfieldly}.  \bg\ decays give electron recoils (ER), and, as we have noted, in \lar\ the main ER/NR discriminant is the PSD available in the time structure of the \sone\ pulse, with additional ER rejection available in \stwo/\sone.
In this analysis, only PSD is used.  The \fno\ cut and minimum and maximum energies define the WIMP search region in the \sone-\fno\ plane, as discussed in Sec.~\ref{sec:wimp_search}.

Neutron scattering in \lar\ gives nuclear recoils, and is thus a more pernicious background.  Our inventory of radioactivity in the detector components indicates that the largest source of internal neutrons is the \tpc\ {\pmt s}, and Monte Carlo calculations matched to the observed conditions in Hall C indicate that internal radiogenic neutrons are more common than cosmogenic neutrons that penetrate the vetoes undetected, with the latter expected to be \dsfbckneucos\ event in a multi-year \dsf\ exposure~\cite{empl}.  
We can distinguish neutrons from WIMPs by observing a coincident event (prompt or delayed) in the \lsv. The TPC adds to this rejection, with $>$1 \stwo\ pulse in an event indicating multiple interactions that will not be present in a WIMP-induced event.
We use trace radioactivity measurements of three early-production samples of the \dsfpmt\ PMT, the $(\alpha,n)$ yield in the PMT materials, and 
Monte Carlo studies of neutron-induced nuclear recoil events in the TPC to estimate the expected number of neutron events.  For the exposure reported here, we expect \dsfruntimeneuexp\ neutron events, with large ($>$ factor of 2) uncertainties, in the WIMP search region after all TPC cuts but before neutron veto cuts.

Surface backgrounds come from the $\alpha$ decays of radon daughters or other trace radioactivity on or just under surfaces in contact with the active \lar.  Both the daughter nuclides and the alphas from these decays are NR background. Typical alphas, starting at energies well above that expected for NR from WIMP scattering, are only a problem if they start deeper beneath the surface.  In \dsf, all surfaces in contact with the \lar\ except the grid are coated with \tpb.
In both decays on the \tpb\ surface with daughter nuclei entering the \lar\ and decays beneath the \tpb\  sending an $\alpha$ into the \lar, the $\alpha$ deposits energy in the \tpb, giving a scintillation signal that mixes with \sone~\cite{deap-pollmann}.
The \tpb\ was evaporated onto the surfaces in a radon-suppressed clean room, and the coated parts remained in a radon-suppressed environment from then on.  Fiducialization can be applied to remove potential surface background that remains.

The ER rejection needed to deal with \ar\ is so high
in atmospheric-argon based WIMP searches that surviving backgrounds from \gr s and other $\beta$ decays, notably \bg-emitting impurities in the LAr, are completely negligible.
An exception is multiple-sited events with one ER in the \lar\ and a second in a transparent material in the TPC, most notably the fused-silica anode, cathode, and \pmt\ windows.  The underlying event can be multiple Compton scattering of a \gr\ or a correlated $\beta+\gamma$ from radioactive decay(s).  Prompt Cherenkov radiation of the recoiling electron in the transparent material mixes with the \sone\ emission from the \lar, increasing \fno\ and, as there is no ionization collected from one of the ER, decreasing \stwo/\sone.  The major tool for dealing with such background from the fused silica is the fraction of \sone\ light in a single \pmt, as the Cherenkov light will usually be emitted in, or directly in front of, one \pmt. 

The present data set was acquired between November~2013 and May~2014.  
The data set is divided into runs of typically \num{200000} or \num{400000} triggered events each, lasting up to about 8 hours.  The usable livetime, defined as all runs taken in dark-matter search mode with a drift field of \dsfdriftfield\ and with all three detectors included, was \dgcritliv.

\begin{table*}[t!]
\begin{tabular*}{1.1\textwidth}{@{\extracolsep{\fill}}llSSSS}
\hline\hline
&{\bf Cut}
						&{\bf Residual Livetime}
									&{\bf Acceptance}
												&{\bf Fiducial Mass}\\
\hline
\parbox[t]{3mm}{\multirow{3}{*}{\rotatebox[origin=c]{90}{\bf Run}}}
&Usable runs				&\dgcritliv 		&			&			\\
&Automated selection		&\dccritliv 		&			&			\\
&Single run				&\nccutliv 		&			&			\\
\hline
\parbox[t]{3mm}{\multirow{4}{*}{\rotatebox[origin=c]{90}{\bf Quality}}}
&Baseline found			&\bacutliv		&			&			\\
&Time since previous trigger	&\licutliv		&			&			\\
&Large gap				&\lwcutliv		&			&			\\
&Veto data present			&\vpcutliv		&			&			\\
\hline
\parbox[t]{3mm}{\multirow{9}{*}{\rotatebox[origin=c]{90}{\bf Physics}}}
&Number of pulses			&			&\npcutacc	&			\\
&First pulse time			&			&\ttcutacc		&			\\
&No \sone\ saturation		&			&\nscutacc	&			\\
&\stwo\ pulse shape			&			&\vwcutacc	&			\\
&Minimum \stwo			&			&\uwcutacc	&			\\
&Max \sone\ fraction per \pmt	&			&\mfcutacc	&			\\
&Prompt \lsv				&			&\cvcutacc	&			\\
&Delayed \lsv\ and \wcd		&			&\hvcutacc	&			\\
&Drift time fiducialization		&			&			&\dfcutfid		\\
\hline
&Total					&\totalliv		&\totalacc		&\dfcutfid		\\
\hline\hline
\end{tabular*}
\caption{List of cuts and their effects on livetime, acceptance, and fiducial volume.  
Where quoted, the errors are systematic.  We do not quote the statistical errors, which are negligibly small.}
\label{tab:acceptance}
\end{table*}

We set criteria for removing runs based on information automatically stored in the run database.  We remove runs that were very short (which indicated DAQ problems), runs with inconsistent livetime, inhibit time, and elapsed time, and runs with an abnormally low live fraction.  Finally, we eliminate \num{37} runs manually based on logbook entries that indicated, for example, that a single \pmt\ was off or that the run was not to be used.  The resulting livetime after applying the run selection criteria is \srcritliv, shown in Table~\ref{tab:acceptance}.

Event-by-event data quality cuts are applied, eliminating events in which the baseline-finder failed on any TPC channel, there was no GPS-timestamp-matched veto data, or there was a $>$\lwcutmax\ period since the previous trigger.  The last two of these were due to occasional DAQ problems, the veto problem usually due to the veto DAQ ending its run before the TPC DAQ and the large gap between events leading to a suspect livetime.  Events that occurred less than \licutmin\ after the previous trigger were cut to eliminate events whose \sone\ might have occurred during the deadtime of the earlier event.  This condition effectively reduced the livetime prior to the surviving events as well.
The data quality cuts together cost about \SI{3}{\percent} of the exposure, summarized in Table~\ref{tab:acceptance}.

The uncertainty on the total livetime for surviving events does not include statistical uncertainties, as these are small.  The systematic uncertainty is dominated by the accuracy of the trigger board timer that measures the livetime, which has been verified to the level of \SI{0.5}{\percent}.  Systematic uncertainty arising from the changing of the DAQ acquisition window or from effects on the livetime definition due to different triggering conditions are negligible with respect to the dominant uncertainty.

We performed a non-blind physics analysis on the surviving data set.  The acceptance of each cut, shown in Table~\ref{tab:acceptance}, is checked using a combination of \scene\ data~\cite{scene1,scene2}, Monte Carlo simulation, and data from \dsf\ itself.  We impose several classes of cuts: 

\begin{enumerate}

\item Selection of single-sited events in the \tpc, eliminating some neutron- and \gr-induced background, begins by requiring that events contain two pulses, allowing a third if its timing with respect to the second is consistent with an \sthree.  The acceptance of this cut is evaluated by examining rejected events and individually studying and accounting for many cases (number of pulses and types of pulses) to determine the fraction of single-sited events that might fall into each category.  The acceptance is \npcutacc, with most of the loss due to accidentals or to events in which the pulse-finder identified one or more extra pulses, usually in the tail of S2. Losses due to inefficiency of the reconstruction algorithms in identifying S1 and S2 pulses are negligible in the WIMP search region. 

\item Cuts to establish the validity of the \sone-\stwo\ identification of the found pulses are applied to allow use of these pulses for PSD and fiducialization.  We require that the first pulse occur at the expected time in the acquisition window to within \ttcuthalfwindow, consistent with our assumption that we triggered on \sone.  This cuts many classes of events, including ``junk" events like triggers on the tails of previous events.  The acceptance loss for real WIMP scatters would be from accidentals. This is evaluated from the measured loss of events by correcting for the non-accidental fraction by hand-scanning. 

We require that the \sone\ pulse not saturate the electronics.  Studying the effect on the \ar\ spectrum indicates that the acceptance is essentially unity in our WIMP search region.  Note that we do not apply the saturation cut to \stwo\ in this analysis.

To confirm the identity of the second pulse as \stwo, we check its pulse shape, using \fno\ of the {\it second\/} pulse and requiring that it be less than \vwcutmax. Because the rise time of \stwo\ is $\sim$\SI{1}{\us}, its \fno\ is typically $<$\num{0.05}, and the acceptance of this cut is essentially unity.  We also require that the \stwo\ pulse be larger than \uwcutmin, where a typical value is $>$\SI{1000}{\pe} for events at the lowest energies used in this analysis.  The acceptance for this cut is estimated using SCENE 36 keV NR data \cite{scene1}, correcting for the relative light yields of the two detectors and taking into account the radial dependence of \stwo\ observed in \dsf.  

\item We see evidence for Cherenkov background, including a sample of events with both \fno$\approx$\num{1} and nearly all the \sone\ signal in a single \pmt.  We cut events in which the \sone\ light is abnormally concentrated in a single \pmt.  The cut is \sone-dependent through the fluctuation statistics, and is position dependent, as events near the top and bottom of the \tpc\ naturally have their scintillation light concentrated more on a single \pmt.  The cut is designed to retain 99\% of events in each bin in the drift time vs. \sone\ plane.  Monte Carlo studies suggest that the vast majority of such background events will result in a sufficient concentration of light in a single PMT to allow rejection by this approach. 

\item Veto detector information is used to suppress events with either prompt energy deposition in the \lsv\ from neutron thermalization, or delayed energy deposition from neutron-capture \grs, notably those from the dominant \lithexc\ final state from capture on \borten\ in the scintillator.  The prompt region of interest (ROI) is especially important in this analysis, as the high rates from \cfor\ in this data preclude thresholds low enough to veto captures to the \lith\ ground state.  The prompt ROI is defined as \cvcutwin\ relative to the observed time of coincidences with \tpc\ events.  Its duration is based on the light collection time in the \lsv, the neutron time-of-flight, and the thermalization time.  Events with more than \cvcutpemin\ ($\sim$\cvcutenemin) in the \lsv\ prompt ROI are vetoed.

Two delayed ROIs are defined in the \lsv.  The first is the \hvcutsliderwidth\ window with the maximum observed charge in the interval from prompt to \hvcutdelwin.  Events with more than \hvcutdelpemax\ ($\sim$\hvcutdelenemax) in this ROI are vetoed.  This ROI covers four neutron capture lifetimes in the borated scintillator and its threshold is chosen above the bulk of the observed \cfor\ signal.  The second delayed ROI is the \hvcutsliderwidth\ window with the maximum observed charge in the interval between \hvcutdelwin\ and the end of the LSV acquisition gate.  Events with more than \hvcutlatepemax\ ($\sim$\hvcutlateenemax) in this ROI are vetoed.  This cut is intended to catch neutrons that thermalize in detector components with long capture lifetimes.  Finally, events are cut if they have more than \hvcutlatectfpemax\ recorded in the entire acquisition window of the \wcd.

The rejection of single-sited neutron-induced events in the TPC WIMP search region by the \lsv\ cuts is estimated with Monte Carlo to be a factor of about \odneutronrejectionachieved, which corresponds to a neutron detection efficiency of $\sim$0.98.  The acceptance loss due to accidentals is determined by counting events rejected by the veto cuts, with the prompt ROI replaced by one of the same duration \SI{2}{\micro\second} before the prompt time.  The accidental acceptance loss of all the veto cuts together is 11\%.  The error is statistical and negligible.

\item The fiducial volume is limited in the vertical coordinate (measured by electron drift time) only -- no radial cut is applied.  Signal-like (high \fno) events are observed near the grid and cathode, and their origin is under investigation.  
We place a fiducial cut retaining events with drift times between \dfcutmin\ and \dfcutmax, corresponding to \dfcutzmin\ below the grid and \dfcutzmin\ above the cathode.  This reduces the total active volume to \dfcutfid, where the dominant uncertainty arises from the uncertainty on the shrinkage of the teflon body of the TPC when cooled from room temperature to cryogenic temperature.

The lowest-achieved level of surface contamination by alpha emitters is $<$10~$\alpha$'s/(m$^2$-d)~\cite{pocar,bx-vessels,sno-ncd}.  Even at this level, we would expect to observe surface events from the TPB-coated cylindrical reflector, with additional contribution to the light signals from the TPB's own scintillation~\cite{deap-pollmann}.  There is no such background left in the WIMP search region after all TPC cuts.   Preliminary studies suggest that $x$-$y$ reconstruction of events in \dsf\ should allow radial fiducialization to suppress any surface background that may become evident in longer running. 

\end{enumerate}

Table~\ref{tab:acceptance} shows the effect of each cut on either the residual livetime, the acceptance for nuclear recoils, or the fiducial mass, along with the estimated systematic uncertainty of each.   The final cuts, on minimum and maximum \sone\ and \fno, define the WIMP search region and are discussed below.
\section{WIMP Search}
\label{sec:wimp_search}

\begin{figure*}[!t]
\includegraphics[width=\textwidth]{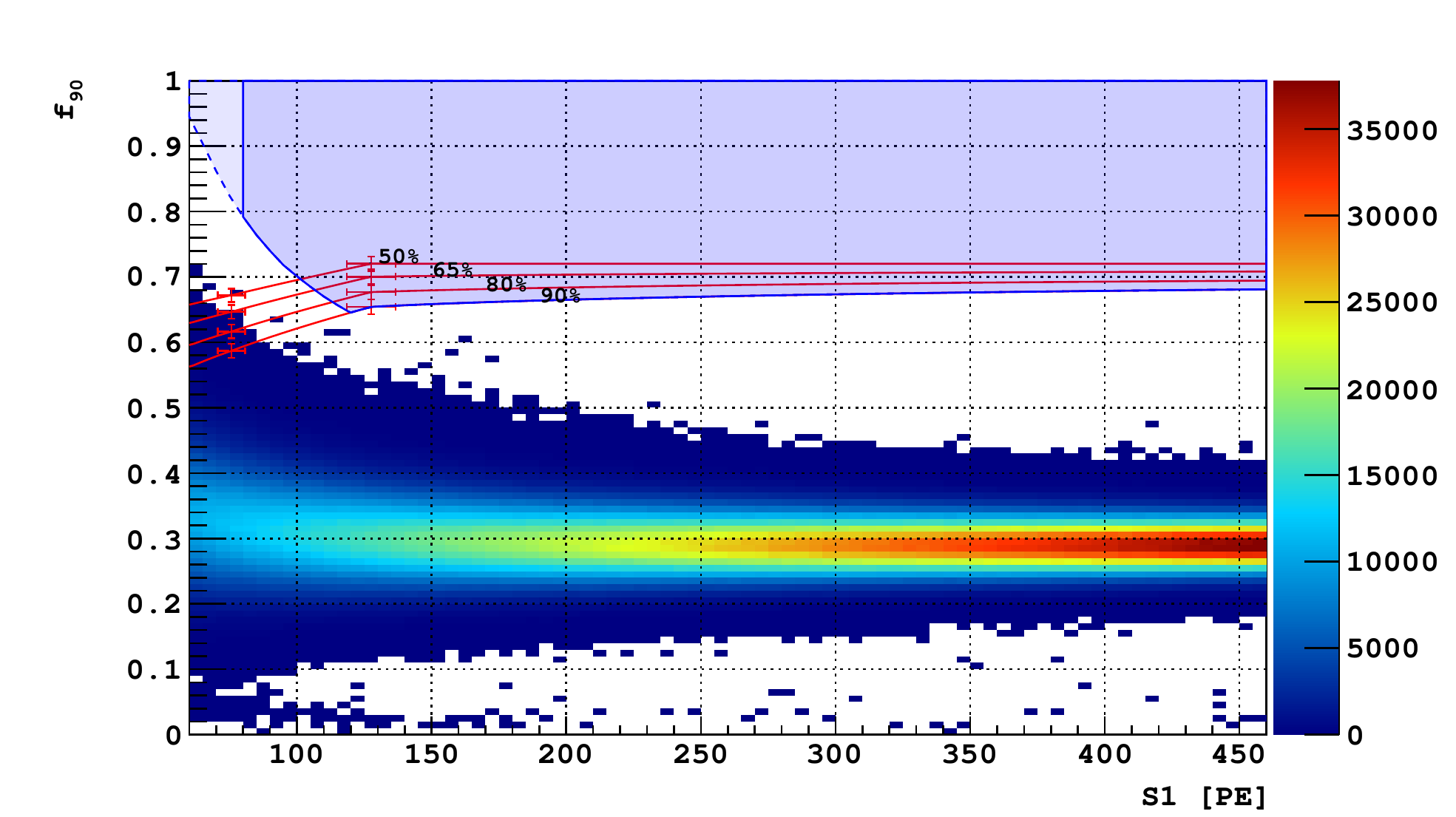}
\caption{Distribution of the events in the scatter plot of \sone\ vs. \fno\ after all quality and physics cuts.  Shaded blue with solid blue outline: dark matter search box in the \fno\ vs. \sone\ plane.  Percentages label the \fno\ acceptance contours for nuclear recoils drawn connecting points (shown with error bars) determined from the corresponding SCENE measurements.
}
\label{fig:dms}
\end{figure*}

The total exposure (fiducial volume $\times$ livetime $\times$ acceptance) remaining after all cuts prior to the WIMP search box is \dsfexpo.  The distribution of the remaining events in the scatter plot of  \fno\ vs.~\sone\ after all quality and physics cuts is shown in Fig.~\ref{fig:dms}.  
There are \dsfnumareventsinplot\ events in this plot, dominated by \ar\ decays.

\begin{figure*}[!t]
\includegraphics[width=0.49\textwidth]{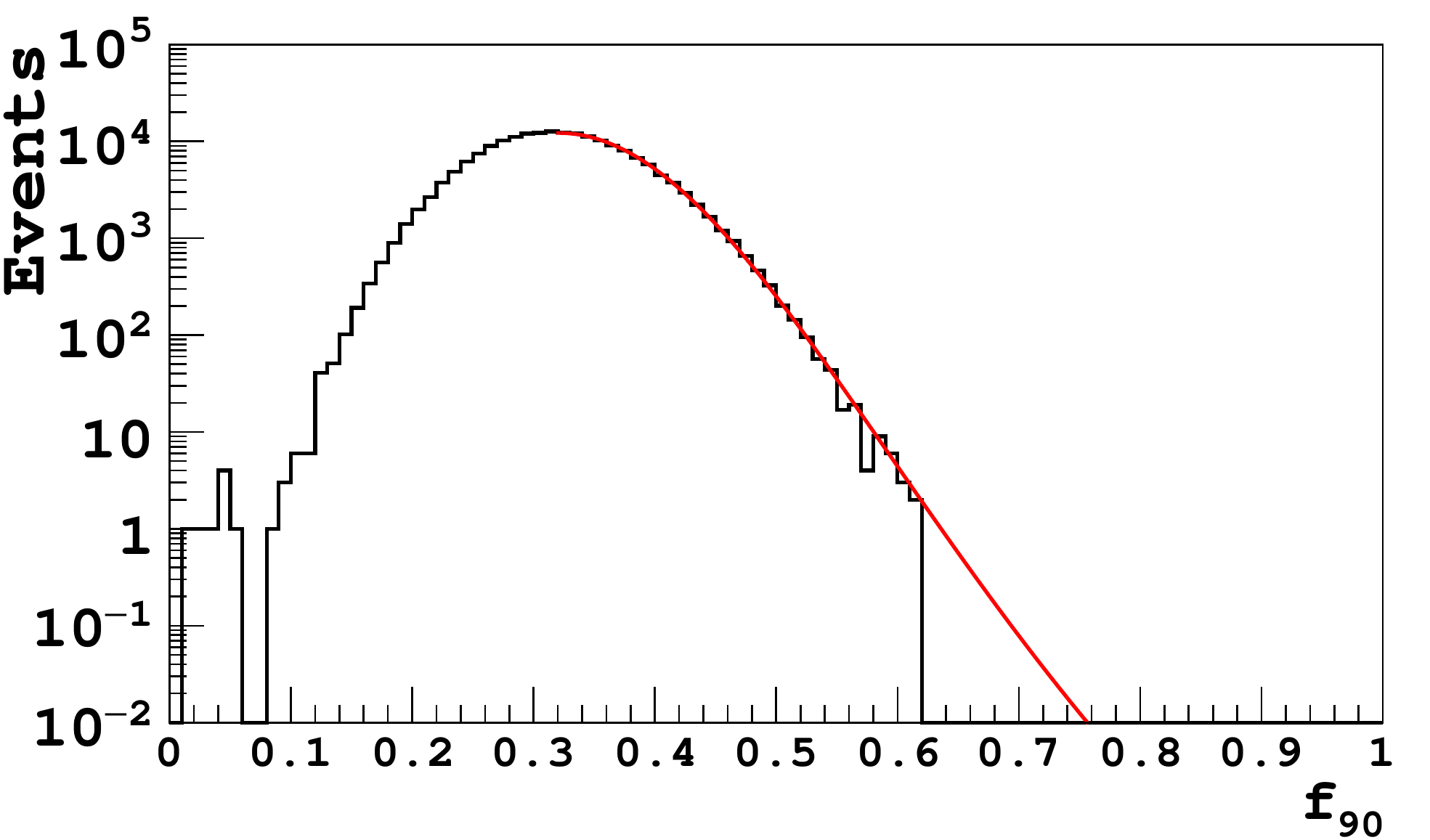}
\includegraphics[width=0.49\textwidth]{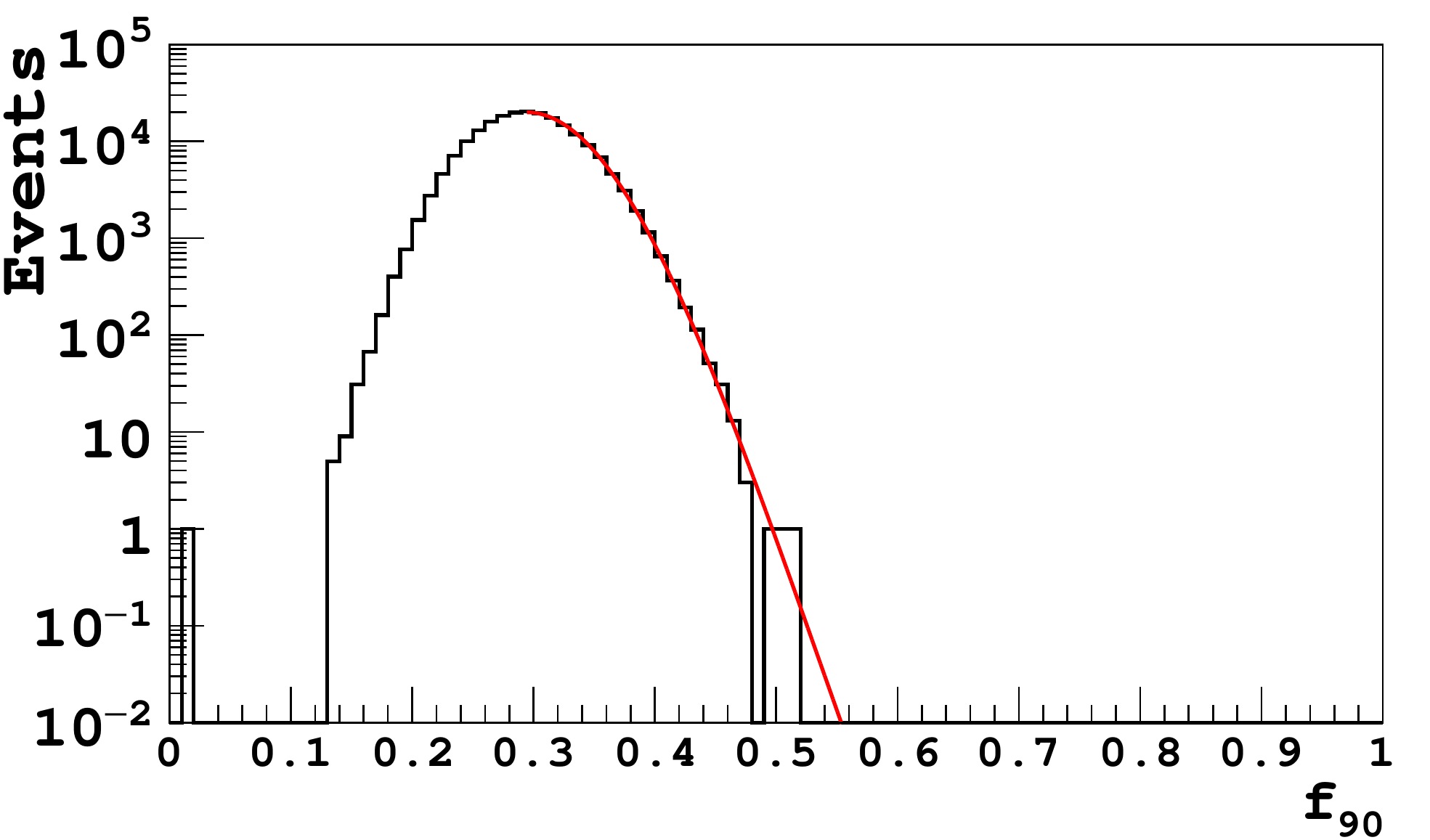}
\caption{Fits of \fno\ experimental distributions using the \fno\ model introduced in Ref.~\cite{lippincott,deap}.  Left: fit for the lowest bin in the WIMP search region, \SIrange{80}{85}{\pe}. Right: fit for a typical higher-energy bin, \SIrange{180}{185}{\pe}.}
\label{fig:f90fits}
\end{figure*}

This distribution was studied by dividing the events into \SI{5}{\pe}-wide slices in \sone\ and fitting the resulting distributions with an approximate, analytical statistical model of \fno\ introduced in Ref.~\cite{lippincott} 
and used in Ref.~\cite{deap} to characterize the \fno\ distribution in \lar\ of a large statistics (\num{1.7E7}) sample of \gr-scatters.  
The important parts of the model are the contributions to the variances of the prompt and late charges (in \si{\pe}) that determine \fno.  
The largest contributions are from the photoelectron Poisson statistics, given by the mean charges themselves.
The variance of the SPE charge distribution itself is also known -- it is determined as part of the SPE calibration.  
The remaining variance is parametrized empirically by two terms: a term proportional to the charge that applies to both the prompt and late charges and, for the late charge, a constant term to represent contributions including electronic noise.
(The variance of the prompt charge due to electronic noise is found to be negligible.)
With the measured variance of the \fno\ distribution in each slice used to constrain the constant term in terms of the other contributions, the only remaining unknown in the variance is the empirical term proportional to charge.
The model is then fit to each slice with the fraction of prompt light (median \fno), the unknown empirical factor, and an overall normalization factor the only fit parameters. 
The empirical factor is found to be the same for all bins.  Measured \fno\ distributions and fits are shown for the lowest bin in the WIMP search region (defined below) and a typical high energy bin in Fig.~\ref{fig:f90fits}.  The model generally provides a good match to the tails of the experimental \fno\ distributions above \SI{120}{\pe}, while below this value the model overestimates the tails. 

Nuclear recoil acceptance curves in the \fno\ vs. \sone\ plane are derived from \scene\ \fno\ medians.  These \fno\ median values from \scene, linearly interpolated and assumed to be constant above the highest SCENE NR energy, are translated from true nuclear recoil energy
to \dsf\ \sone\ values using the information in Tables~\ref{tab:ly} and \ref{tab:scene}.   This gives the 50\% contour for \dsf.  
The other contours and associated errors depend also on the width of the \dsf\ \fno\ distribution at each \sone.
For this we use the same analytical \fno\ model described above.  Aside from the \fno\ median at each \sone, all the other parameters in the model remain fixed from the fits to the high-statistics \ar\ data at the same \sone.
The resulting acceptance curves are shown in Fig.~\ref{fig:dms}.

\begin{figure}[!t]
\includegraphics[width=\columnwidth]{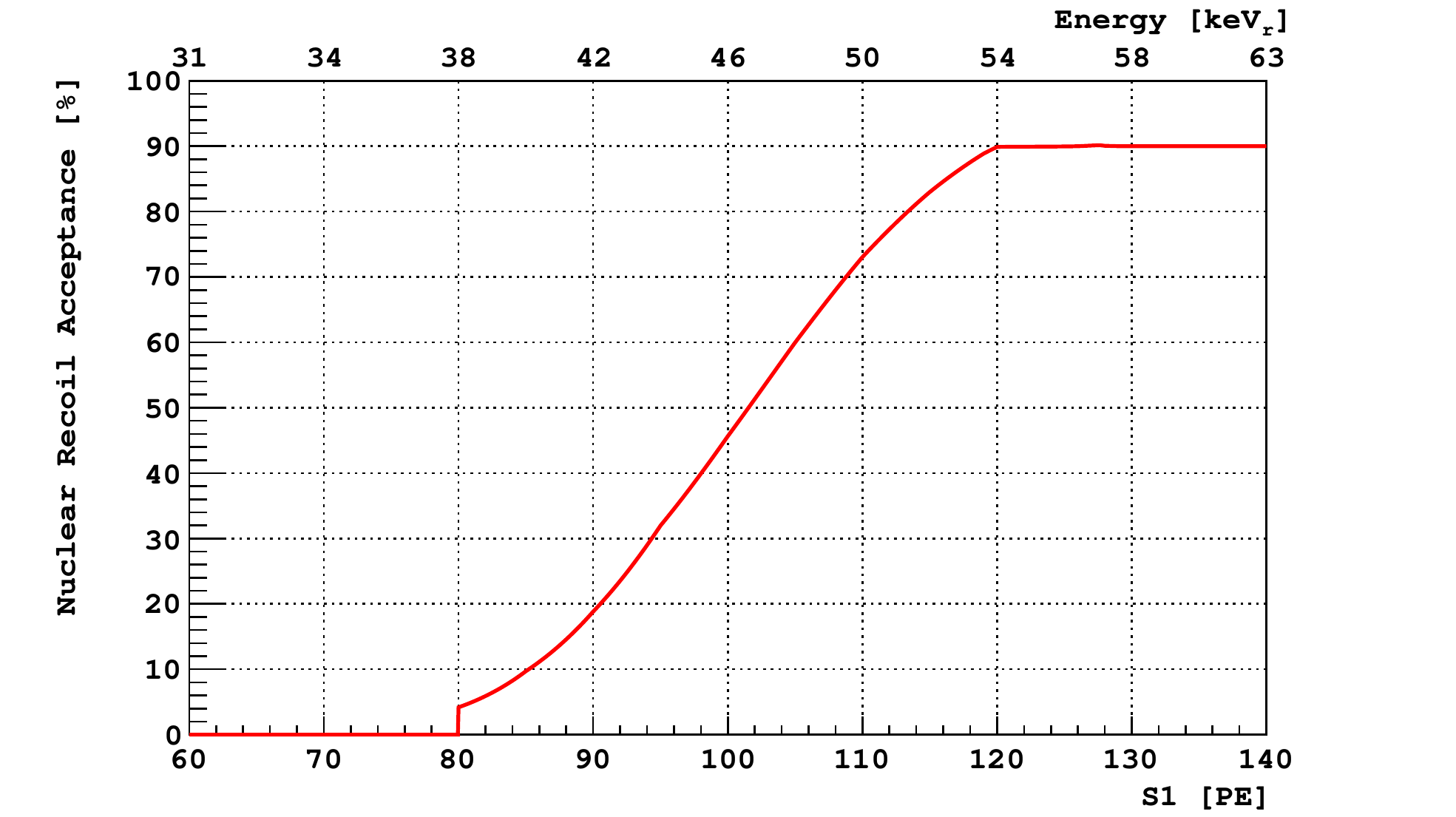}
\caption{Nuclear recoil acceptance of the dark matter search box. Acceptance is fixed at $90\%$ between 120 and 460 PE (54 and \SI{206}{\keVr}).
}
\label{fig:accs1}
\end{figure}

The dark matter search box shown in Fig.~\ref{fig:dms} is obtained by intersecting the \SI{90}{\percent} nuclear recoil acceptance line with the curve corresponding to a leakage of \ar\ events of \isovalgreen\ according to the statistical model for electron-recoil \fno\ described above.  This bound leads to an expected leakage of \ar\ into the full search box, bounded by \socutmin$<$\sone$<$\socutmaxpaper\ (\socutenemin$<$$E_{\rm recoil}$$<$\socutenemaxpaper), of $<$\dsfbckargreen.  
The lower bound in \sone\ is chosen where the acceptance for WIMPs above the leakage curve drops below \SI{5}{\percent} (see Fig.~\ref{fig:accs1}), 
while the upper bound is chosen to contain most of the integrated acceptance for WIMPs in the standard halo model discussed below.
There are no events in the search region.

We observe \dsfruntimeneuobs\ events passing all \tpc\ cuts and with nuclear-recoil-like \fno, but with energy depositions in the \lsv\ above our veto cut threshold.  In coincidence with one of these \dsfruntimeneuobs\ neutron candidates, we recorded signals near saturation in both the \lsv\ and the \wcd, and therefore we classify that event as cosmogenic, leaving \dsfruntimeneuradobs\ radiogenic neutron candidates.  
This is to be compared to the \dsfruntimeneuexp\ neutron-induced events expected from the Monte Carlo studies of PMT radioactivity discussed in Section~\ref{sec:analysis}.

\begin{figure}[!t]
\includegraphics[width=\columnwidth]{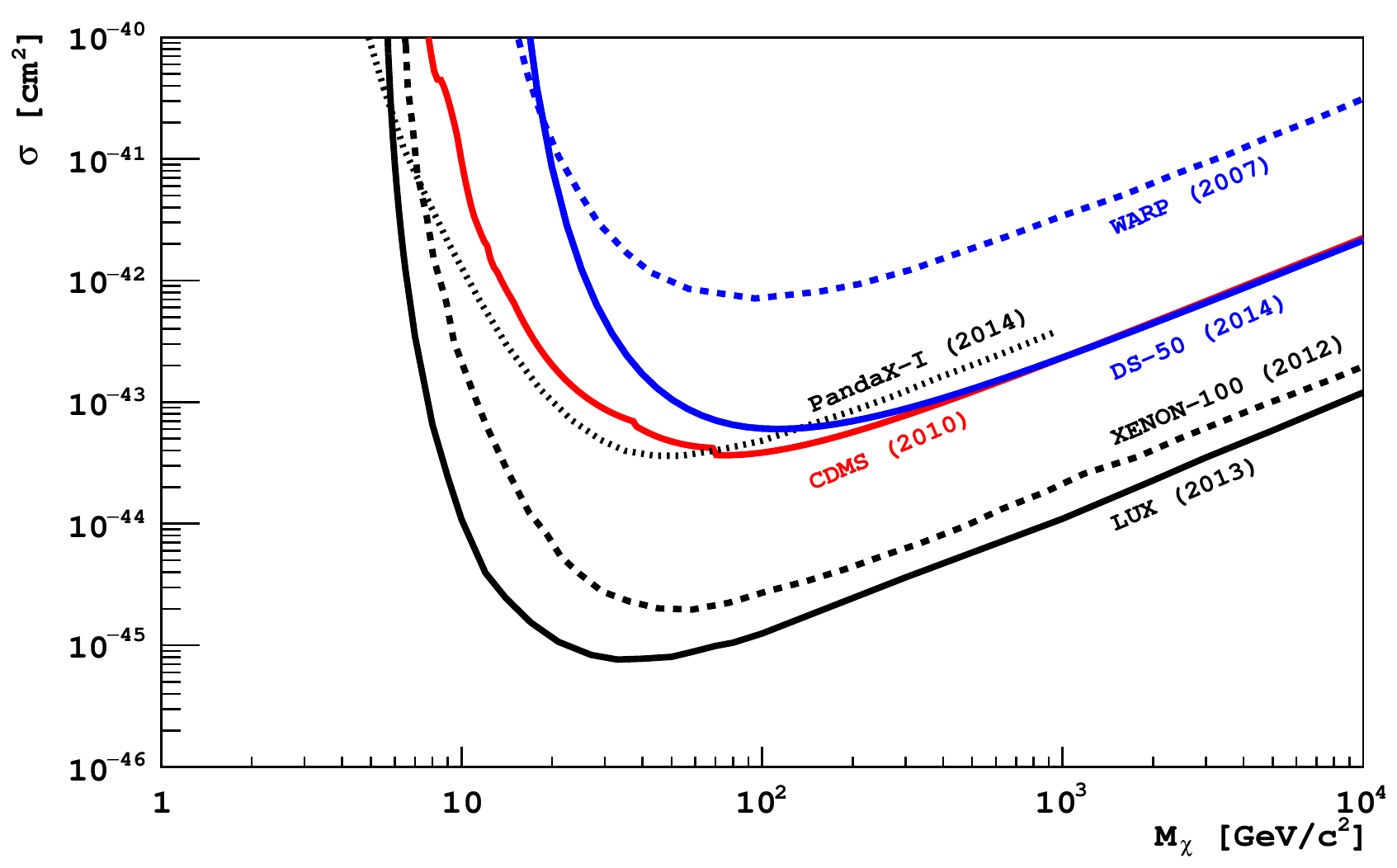}
\caption{Spin-independent WIMP-nucleon cross section \SI{90}{\percent}~C.L. exclusion plot for the \dsf\ atmospheric argon campaign (solid blue) compared with results from LUX~\cite{lux} (solid black), XENON100~\cite{xenon-100} (dashed black), PandaX~\cite{pandax-i} (dotted black), CDMS~\cite{cdms} (solid red), and WARP~\cite{warp} (dashed blue).  
}
\label{fig:sensitivity}
\end{figure}

To derive a dark matter limit from Fig.~\ref{fig:dms}, we assume the standard isothermal-WIMP-halo model~\cite{lewin,savage} with \vesc=\SI{544}{\km\per\s}~\cite{smith}, \vnaught=\SI{220}{\km\per\s}~\cite{smith},\linebreak\ 
\vearth=\SI{232}{\km\per\s}~\cite{gelmini}, \rhodm=\SI{0.3}{\GeV\per\square\c\per\cubic\cm}~\cite{savage}.  Given the null result shown in Fig.~\ref{fig:dms}, we derive a \SI{90}{\percent} C.L. exclusion curve  corresponding to the observation of \SI{2.3}{\ev} for spin-independent interactions, and we compare it in Fig.~\ref{fig:sensitivity} with limits from recent experiments.

\section{Conclusions}
\label{sec:conclusions}

We report on the first underground operations for physics data taking using the complete \dsf\ direct dark matter search detection system, including the \lar\ \tpc, the liquid scintillator shield/veto, and the water-Cherenkov shield/veto.  An innovative closed-loop argon circulation system with external purification and cooling allows the \lar\ \tpc\ to achieve an electron drift lifetime of \dsfelectronmeanlifesecond.  Photoelectron yield of \dsfnullfieldly\ at null field is achieved for detection of the primary argon scintillation, giving the photoelectron statistics necessary for high performance pulse shape discrimination.

Figure~\ref{fig:dms} covers the range of energies from \SIrange{8.6}{65.6}{\keV} for \ar, and a total of \dsfnumareventsinplot\ \ar\ events were recorded over that energy range.  
Event selection based on the \tpc\ cuts is shown to completely suppress \ar\ background events in the present \dsfexpo\ exposure.

This exposure contains at least as many \ar\ events as \dsfuareqexponoerr, or \dsfuareqexpoty, of running with \uar, proving that \dsf\ could run for two decades with \uar\ and be free of \ar\ background.  Alternatively, we note that the WIMP search region in even the longest contemplated \dsf\ \uar\ run, drawn to admit the same \isovalgreen\ of \ar\ as the analysis reported here, would move lower in \fno, giving higher WIMP acceptance at low energies.

Although the liquid scintillator veto was compromised by a high \cfor\ content during this exposure, it was able to tag and remove the handful of neutron events expected.  In the \uar\ run, we will be operating with a neutron veto that will be able to sustain lower thresholds, predicted to give considerably higher neutron rejection factor.

A WIMP search with the present dataset gives a limit as low as \dsflimit\ at 100 GeV/$c^2$, the best result achieved to date with an argon target.
\section{Acknowledgments}
\label{sec:acknowledgments}

We thank Christopher Condon, Robert Klemmer, Matthew Komor, and Michael Souza for their technical contributions to DarkSide.
We thank Matthias Laubenstein of LNGS for his numerous radiopurity measurements of \dsf\ components.
We acknowledge support from the NSF (US, Grants PHY-0919363, PHY-1004072, PHY-1211308, and associated collaborative Grants), DOE (US, Contract Nos.~DE-FG02-91ER40671 and DE-AC02-07CH11359), the Istituto Nazionale di Fisica Nucleare (Italy), and the NCN (Poland, Grant UMO-2012/05/E/ST2/02333).
This work was supported in part by the Kavli Institute for Cosmological Physics at the University of Chicago through grant NSF PHY-1125897 and an endowment from the Kavli Foundation and its founder Fred Kavli.





\end{document}